\documentclass[prx,twocolumn]{revtex4-2}
\usepackage{graphicx}
\usepackage{amsmath}
\usepackage{natbib}
\usepackage{epstopdf}
\usepackage{xcolor}

\newcommand{\bn}{\begin{eqnarray}}
\newcommand{\en}{\end{eqnarray}}
\newcommand{\eml}{\end{multline}}
\newcommand{\bml}{\begin{multline}}
\newcommand{\h}{\hspace}

\begin{document}

\title {Quantum States and Spectra of Small Cylindrical and Toroidal Lattices}

 \author{Caelan Brooks$^{1,2}$ and Kunal K. Das$^{2,3}$}
  \affiliation{$^1$Department of Physics, Harvard University, Cambridge, Massachusetts 02138, USA}
 \affiliation{$^2$Department of Physical Sciences, Kutztown University of Pennsylvania, Kutztown, Pennsylvania 19530, USA}
\affiliation{$^3$Department of Physics and Astronomy, Stony Brook University, New York 11794-3800, USA}
\date{\today }

\begin{abstract}
We examine the spectrum and quantum states of small lattices with cylindrical and toroidal topology subject to a scalar gauge potential that introduces a position dependent phase in the inter-site coupling. Equivalency of gauges assumed in infinite lattices is generally lost due to the periodic boundary conditions, and conditions that restore it are identified. We trace the impact of various system parameters including gauge choice, boundary conditions and inter-site coupling strengths, and an additional axial field. We find gauge dependent appearance of avoided crossings and persistent degeneracies, and we show their impact on the associated eigenstates. Smaller lattices develop prominent gaps  in spectral lines associated with edge states, which are suppressed in the thermodynamic limit. Toroidal lattices have counterparts of  most of the features observed in cylindrical lattices, but notably they display a transition from localization to delocalization determined by the relation between the field parameter and the number of lattice sites.
\end{abstract}
\maketitle

\section{Introduction}

The quantum mechanical behavior of charged particles in a two dimensional (2D) lattice potential in a magnetic field is an exceptionally rich system, largely responsible for making concepts of topology an essential feature of many body quantum systems \cite{TKNN,Laughlin-PRB-QHE}. The characteristic fractal spectrum  known as the Hofstadter butterfly \cite{Hofstadter-PRB} is one of the most familiar patterns in all of physics. Within the vast literature on this system, studies of its spectral features generally assume an infinite lattice which, in physically relevant cases, can be mapped to a 1D periodic lattice with modulated on-site energies described by the well-known Harper equation \cite{Harper}. Furthermore, in the context of electronic systems, the magnetic field is the fundamental physical entity, and gauge invariance is invoked for a flexible choice of potential.  In this paper, we look beyond these assumptions motivated by broader possibilities in the context of synthetic gauge fields in ultracold atoms \cite{Goldman-Zoller-NaturePhysics,Review-Goldman}.

With synthetic gauge fields, by construction, the vector potential is the fundamental ingredient: the Abelian gauge created via an effective phase on the inter-site coupling \cite{Jaksch-Zoller}.  The flux and the magnetic field are the emergent features.  Non-trivial global topology of the lattices also breaks the equivalence of typically interchangeable gauges \cite{Das_Brooks_short}. Thus, a more bottom up perspective is suggested where the choice of the gauge, such as defined for analogous infinite systems, determines the physical properties. Such synthesized systems with one or more periodic boundaries imposed by the topology, can be scaled down to just a few lattice sites, to accentuate the impact of boundary conditions. That will be our target regime, since smaller systems can also provide insights often lost in the complexity of larger systems. Specifically, small lattices can highlight features of coherent quantum media that are suppressed as the system size increases. Topology can impact quantum states, and not their classical counterparts,  precisely because of the coherence of the former. However, when the physical dimension of a system is large compared to the relevant coherence lengths, some of the more profound impacts of topology are lost. Lattices with a few sites can be made to sustain coherence across the entire system.

Interest in synthetic gauge fields has been driven by the physics of the quantum Hall effect (QHE) \cite{Stone-QHE-book,Thouless-book} and its origins in electronic systems has continued to define the configurations and assumptions invoved in atomic systems. However, optical lattices with ultracold atoms do not always have to be limited by those considerations. Although not natural in electronic systems, cylindrical or toroidal topologies can be constructed using a variety of approaches that have been proposed and some implemented in experiments. Trapping atoms in a ring configuration, topologically equivalent to a cylinder has been utilized in numerous experiments \cite{Campbell_resistive-flow,Phillips_Campbell_superfluid_2013,Phillips_Campbell_hysteresis,K-Wright_ring_PRL}, and azimuthal lattice structure has been demonstrated \cite{Padgett, Zambrini:07}.

 While there have been ideas to construct cylindrical and toroidal configurations in real space \cite{Lacki-cylinder,Lacki-torus,Porto-torus}, experiments in recent years  have taken the route of utilizing synthetic dimensions \cite{Boada-PRL,Spielman-synthetic-dim} to implement such configurations, wherein the periodic boundary conditions is implemented by cyclic coupling of internal states \cite{Zhou_Hall_loc_PRL,Zhou_PRL_cylinder_tori,Spielman-synthetic-cyl,
Zhou_PRX_Hall_cylinder,Fabre_Laughlin_expt,Santoro-synthetic-Hall,Shin_bandgap-closing-PRL,Shin-synthetic-Hall}. Such configurations are leading to confirmation of seminal models as well as the discovery of new phenomena. Considering that only a few internal states are typically involved, of the order of three or four, these systems can be readily adapted to examine the physics of small topological lattices by limiting the number of sites in real space using external confining potentials. Measurement of the system properties can be  done with established imaging methods \cite{A5,A2} supplemented by newer stroboscopic techniques that can access effects of correlations \cite{Lelas_2016}.

\begin{figure}[t]
\centering
\includegraphics[width=\columnwidth]{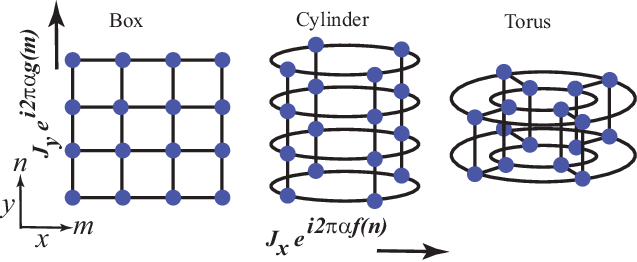}
\caption{(Color online) Schematic of different boundary conditions considered, shown for a $M\times N=4\times 4$ lattice. Box boundary condition is assumed in the non-periodic directions. The inter-site coupling, of strength $J_x,J_y$, contain phase factors $g(m), f(n)$ with site dependence arising from the gauge field, each depending on the complementary index.}
\label{Fig1-Schematic}
\end{figure}

The behavior and dynamics of quantum systems are defined by the nature of the spectrum and the associated eigenstates. Prior studies are characterized by features: Most are focussed on the systems where the spatial boundary conditions are not a significant factor. The few that do consider cylindrical or toroidal
configurations in real space are concerned mostly with the physics associated with the QHE \cite{A10}. This has also had a defining influence for the recent studies with synthetic gauge fields, and specifically the few papers that have started to examine lattices with non-trivial topology in the context of ultracold atoms \cite{Lacki-cylinder,Lacki-torus,Porto-torus, Zhou_PRX_Hall_cylinder,Zhou_PRL_cylinder_tori,Spielman-synthetic-cyl,Shin_bandgap-closing-PRL,Shin-synthetic-Hall}.  While such studies continue to be extremely interesting in their own right, we have a different goal in this paper, which is to understand the impact of the non-trivial topology on the basic spectrum of a 2D lattice subject to an effective magnetic field.

With continuing preoccupation with more esoteric aspects, the more basic features like the spectral properties for non-trivial topologies have not been thoroughly analyzed. Even for trivial topology, the well-known Hofstadter butterfly \cite{Hofstadter-PRB} represents a critical point of that system \cite{Thouless-book}, and hence its most complex incarnation; making it difficult to separate out the impacts of various system parameters. Adding in non-trivial topology in real space complicates it further.  The purpose of this paper is thus to fill in a gap in the literature by providing a comprehensive and comparative study of the spectrum and eigenstates of cylindrical and toroidal lattices as contrasted with planar 2D lattices. We will consider small lattices and take a grounds-up approach by switching on different features, as a general strategy to identify the distinctive features and their influencing factors.  We will also describe the impact of the prominent spectral features on the eigenstates.  This will map out the basic properties of a coherent medium in such topological lattices and help better understand their dynamical behavior of broad interest.

We describe our physical model and assumptions in Sec.~II, and discuss the loss of equivalency of typical gauges in the presence of periodic boundary conditions in Sec.~III along with the conditions under which they are restored. Section IV does a comparative analysis of the general spectral features for a cylindrical lattice, examining the impact of gauge, boundary conditions and inter-site coupling constants, and then the pertinent features of the spectrum are correlated with the behavior of the eigenstates in the Sec. V.  The behavior of edge states in small cylindrical lattices is examined in Sec. VI. We then broaden our analysis to toroidal lattices in Sec. VII and compare how the behavior is further altered by the introduction of a periodic boundary condition in both physical dimensions of a 2D lattice. Our main conclusions and outlook are summarized in Sec.~VIII.

\section{Physical Model}
We consider a quantum mechanical system of particles in a finite 2D lattice potential, its $x$ and $y$ orientations indexed by $m\in\{1,\cdots,M\}$ and $n\in\{1,\cdots,N\}$ respectively and described by a Hamiltonian, $H(n,m)$:
\bn
\sum_{m,n}[J_x e^{-i2\pi\alpha f(n)}\psi_{n,m+1}\h{-1mm}+\h{-1mm}J_ye^{-i2\pi\alpha g(m)} \psi_{n+1,m}+h.c.]\en
where we allow for different nearest neighbor hopping strengths $J_x$ and $J_y$ in the two relevant directions, with lattice spacings $a_x$ and $a_y$. In the presence of a magnetic field in the $z$-direction, we introduce the parameter $\alpha=qBa_xa_y/h$. The factors $f(n)$ and $g(m)$ are then associated with the vector potential and are gauge dependent: for example $f(n)=-n, g(m)=0$ and $f(n)=0, g(m)=m$ would both correspond to the Landau gauge and $f(n)=\mp \frac{1}{2} n, g(m)=\pm \frac{1}{2} m$ would correspond to a symmetric gauge. In the continuum they would represent vector potentials of the form  $\vec{A}=(-By,0,0)$, $\vec{A}=(0,Bx,0)$ and $\vec{A}=(\mp \frac{1}{2}By,\pm \frac{1}{2}Bx,0)$.

Cylindrical topology can be introduced by imposing periodic boundary conditions along either orientation such that $N+1\equiv 1$ or $M+1\equiv 1$, as illustrated in Fig.~1.  Insisting on both creates a torus topology. The Landau gauge choices are particularly suitable for cylindrical symmetry, so our considerations will be framed in terms of them: We will refer to  $f(n)=-n, g(m)=0$ as the Landau $x$ gauge and $f(n)=0, g(m)=m$ as the Landau $y$ gauge; in the rest of the paper, we may simply refer to them as the $x$ gauge and the $y$ gauge respectively.  Switching the signs on the gauge factors reverses the direction of magnetic field.  The periodic boundary conditions  break the equivalency that is typically assumed among these different gauges, when systems are taken to be infinite or have box boundary conditions \cite{Das_Brooks_short}.

\begin{figure}[t]
\centering
\includegraphics[width=\columnwidth]{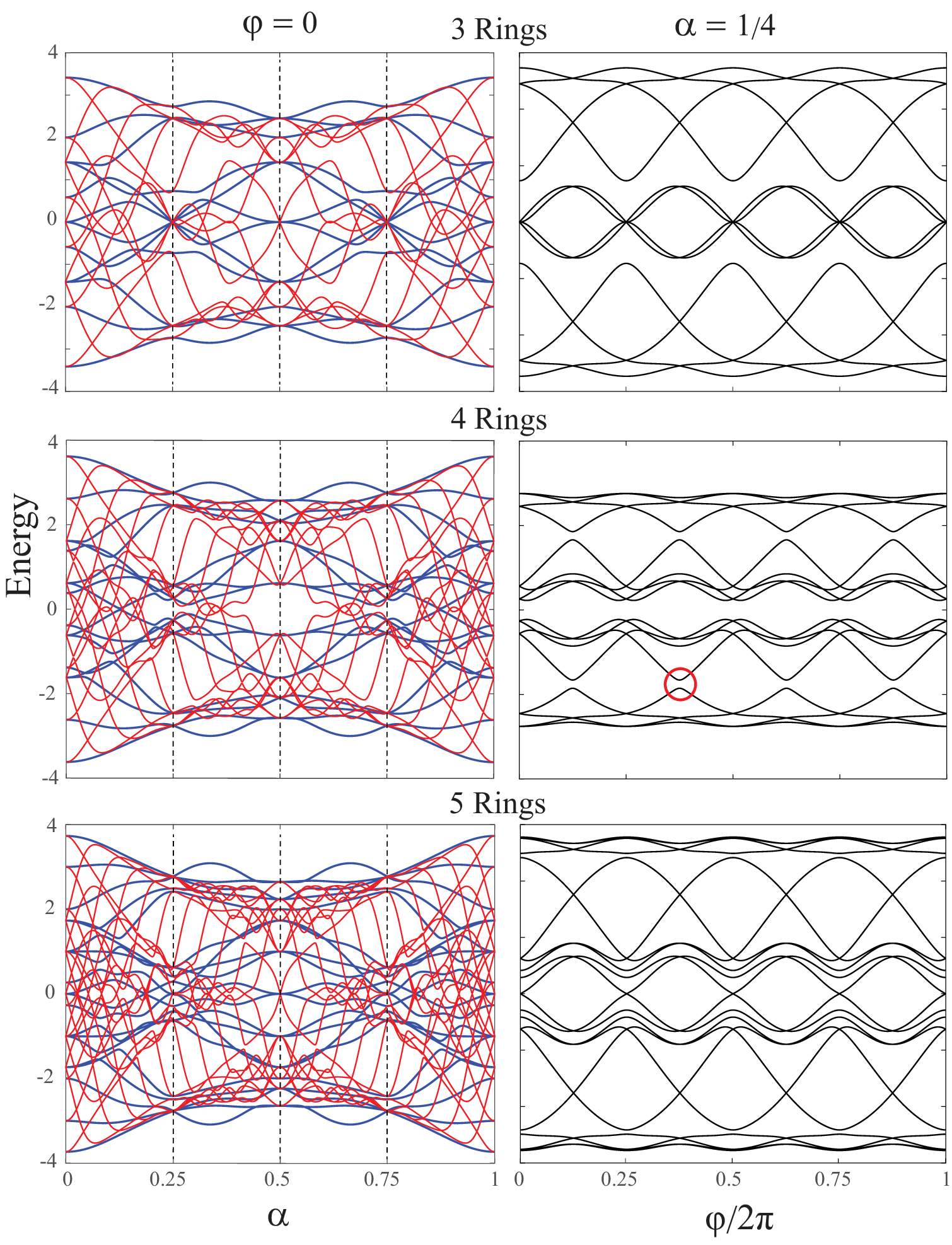}
\caption{(Color online) The periodic boundary condition of a cylindrical lattice breaks the gauge equivalence illustrated for the two Landau gauges $x$ (thin red line)  and $y$ (thick blue line) for a $4\times 4$ cylindrical lattice. The left panels show that they match at specific values of the field where $ M\times\alpha=k$, a natural number, irrespective of  the number of sites $N$ in the axial direction. The right panels illustrate that at those values, here shown for $\alpha=1/4$, the spectrum is the same for both gauges as a function of a phase $\varphi$ representing an added axial field. When $N=M$, gaps open in the lines that mark edge states of the bands; one is marked by a red circle.}
\label{Fig2-xygauge_compare}
\end{figure}

\section{Effect of Gauge on Spectrum}

The breakdown of gauge equivalency in the presence of a periodic boundary condition raises the obvious question: Are there any values of $\alpha$ for which that equivalency is restored? Assuming rational values of $\alpha=p/q$, when the number of sites along the periodic direction is commensurate with the denominator $q$ such that $M\times \alpha=k$, a natural number, then the spectrum in the two Landau gauges coincide. This can be also interpreted as the site number $M$ being commensurate with the magnetic period \cite{Zak-1}. We illustrate this in Fig.~\ref{Fig2-xygauge_compare}, where we superimpose the spectra for the two gauges for three different cylindrical lattices having the same number of sites $M=4$ in the periodic direction, but with different number of parallel rings $N=3,4$ and $5$ that comprise the cylinder. In all cases, the spectra for the two gauges match at $\alpha =1/4,2/4, 3/4, 4/4$, when $M\alpha$ leads to a multiple of $2\pi$ in the total phase change acquired in a circuit of each ring.  The gauge equivalency at those values of $\alpha$ remains when an extra axial field is introduced, like in the well-known Laughlin model for the quantum Hall effect \cite{Laughlin-PRB-QHE}. Such a field presents itself as an added constant phase along the periodic direction $J_x\rightarrow J_x e^{i\varphi}$. At one of those values $\alpha=1/4$ where the eigenvalues for the two Landau gauges coincide in the left panels of Fig.~\ref{Fig2-xygauge_compare}, we also plot the eigenvalues as function of the added phase $\varphi$ in the corresponding right panels.  We find that those patterns remain identical for both Landau gauges.

However, this equivalency does not extend to all choice of gauge, for example the spectrum for the  symmetric gauge differs from the Landau gauges even at those values of $\alpha$ listed above. We can generalize the condition for gauge equivalence that produces a uniform magnetic field in an infinite 2D lattice, of which the Landau and the symmetric gauges are only the simplest possibilities. In general, $\vec{A}=(\mp \frac{r}{s}By,\pm \frac{s-r}{s}Bx,0)$, with $r\in\{0,1,2,\cdots\}, s\in\{1,2,\cdots\}$  would lead to a constant magnetic field $B$ perpendicular to the lattice. In a discrete lattice, this would correspond to $f(n)=\mp \frac{r}{s}n$ and $g(m)=\pm \frac{s-r}{s}m$.  If the lattice is wrapped into a cylinder of $M$ sites in the periodic direction, then $M\times \alpha\times f(1)=k$, a natural number, will lead to invariance with respect to switching the $x$ and $y$ components of the vector potential, and that will remain true for any ratio $J_y/J_x$ allowing for the change in energy scale.

Next, we determine the influence of the number of rings, or site number, $N$  in the non-periodic (axial) direction. As a function of $\alpha$, the spectrum varies substantially with increasing number of sites in either direction. The basic pattern remains self-similar for a specific gauge choice, but gets intricate with more spectral lines.  But, if we consider the spectrum as a function of $\varphi$, the number of sites in the axial direction has a more interesting effect that is apparent in comparing $N=3,4,5$ axial sites in Fig.~\ref{Fig2-xygauge_compare}. When $N$ and $\alpha$ are commensurate, such that $N\alpha=k$ a natural number, increasing $J_y$ opens up gaps in the crossing of the spectral lines that span the gaps between the energy bands; we see those for $N=M=4$ where one is marked by a circle, but not for $N=3$ or $N=5$. This has an impact on the behavior of the states at the band edge as we will discuss later in Sec.~VI.

\begin{figure}[t]
\centering
\includegraphics[width=\columnwidth]{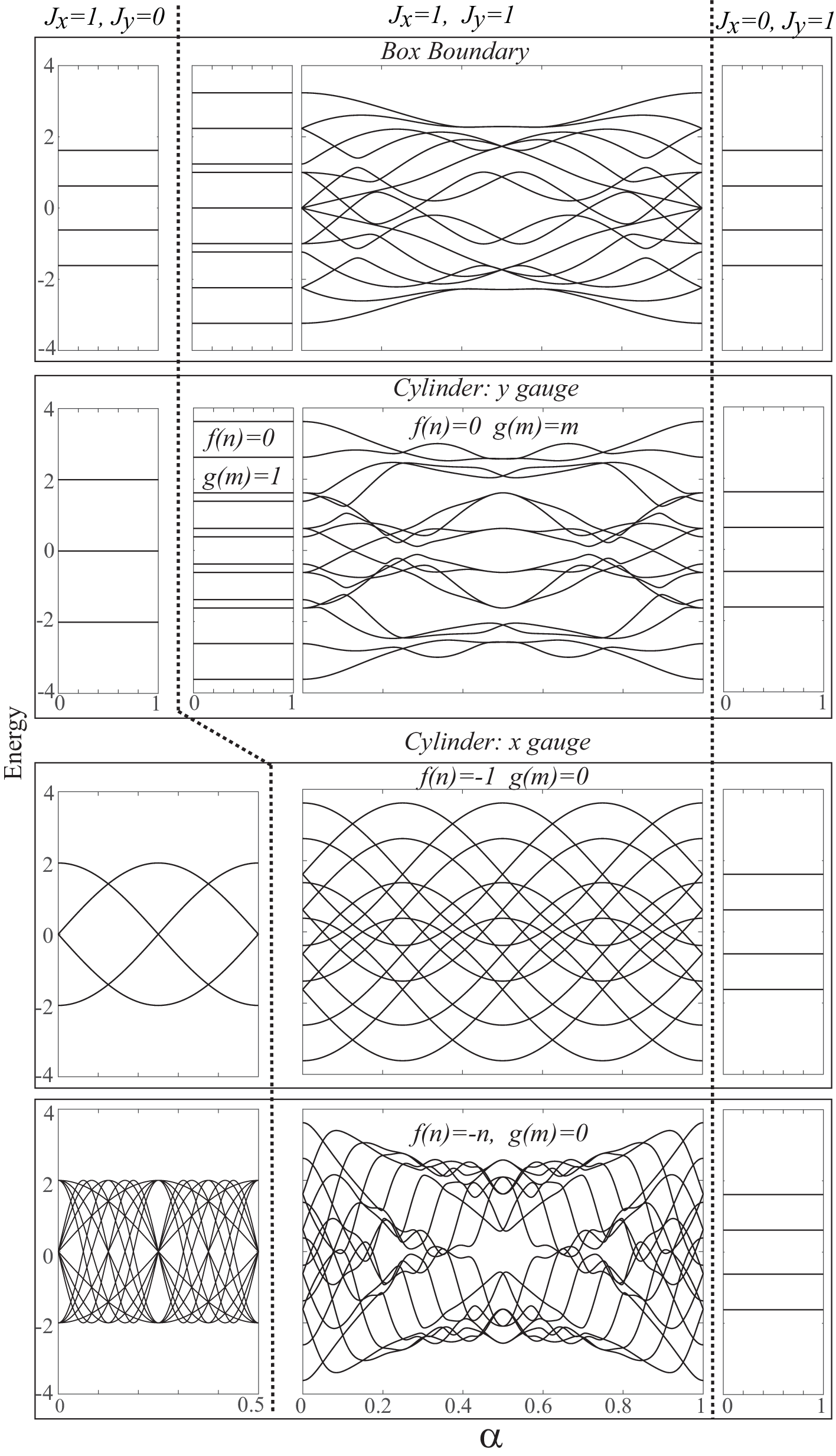}
\caption{(Color online) The spectrum of a $4\times 4$ cylindrical lattice showing effects of boundary conditions, gauge and coupling: The \emph{three columns} separated by dotted lines show: (\emph{Left}) isolated rings $J_x=1,J_y=0$,(\emph{Right}) isolated strips $J_x=0,J_y=1$, and  (\emph{Center}) the critical case $J_x=1,J_y=1$. The \emph{four rows} display: (\emph{First}) \emph{box} boundary condition insensitive to the gauge; (\emph{Second}) Landau $y$ gauge comparing in the center column the effects of $g(m)=1$ with $g(m)=m$, as in the gauge factor being independent/dependent on the lattice index; (\emph{Third}) Landau $x$ gauge showing that the spectrum depends on the field $\alpha$ even when the gauge factor $f(n)=-1$ has no dependence on the lattice index. (\emph{Fourth}) Landau $x$ gauge with that dependence $f(n)=-n$.}
\label{Fig3-Spectralcombo}
\end{figure}

\section{Origins of Spectral features}

The physical behavior of a quantum system is reflected in the spectrum, we will now examine the salient features for a cylindrical lattice with regards to the two different orientations of the Landau gauge. The spectra shown in Fig.~\ref{Fig2-xygauge_compare} are for $J_x=J_y$, which corresponds to the critical case for the Hofstadter butterfly \cite{Hofstadter-PRB}. That case presents the most complex spectral pattern. But, having identical couplings does not carry any significance for cylindrical configuration where the behavior in the  $x$ and $y$ orientations fundamentally differ anyway. We examine the effect of varying the coupling ratio $\gamma=J_y/J_x$. The other degree of freedom ${\textstyle \sqrt{J_x^2+J_y^2}}$ sets the energy scale. The results are summarized in Fig.~\ref{Fig3-Spectralcombo} in the context of an $4\times 4$ lattice with $N$ rings and $M$ azimuthal sites.

It is instructive to examine the limiting cases. When the coupling along the periodic direction vanishes, $J_x=0$, we have $M$ decoupled strips of a 1D lattice, each with box boundary condition and the eigenvalues are
\begin{equation}\label{1Dstrip}
E_k = 2J_y cos\left(\frac{k\pi}{N+1}\right), \h{5mm} k=1,2\cdots N
\end{equation}
These are shown in Fig.~\ref{Fig3-Spectralcombo} on the extreme right panels as $N=4$ four flat lines as a function of $\alpha$, each  $M=4$-fold degenerate, and identical for all the cases shown, even with a $y$ gauge factor along each strip since the box boundary condition makes it physically irrelevant. The opposite limit of vanishing coupling in the open direction $J_y=0$ is more interesting. In this case we have a set of $N$  decoupled rings and the eigenvalues are
\bn \label{1Dring} E_k=2J_x\cos\left(\frac{2\pi m}{M}-2\pi\alpha f(n)\right),\h{4mm} m=1,2\cdots M\en
For the $y$ gauge, phase would then be in the open direction, so the second argument of the cosine vanishes ($f(n) = 0$) and there is no $\alpha$ dependence. Due to a doubly degenerate null eigenvalue for each ring we see three flat lines for this case, as shown on the extreme left of the first two rows in the figure. The effect of $x$ gauge can be understood by first implementing a constant phase $f(n)=-1$ which creates the same dependence on $\alpha$ for all the rings and we get $M=4$ phase-shifted sinusoidal curves, each $N$ fold degenerate. The $x$ gauge $f(n)=-n$ lifts that degeneracy by introducing a different $\alpha$-dependent phase on each ring, creating $N$ intertwined sinusoidal curves with periods $1/n, n=1,2,3,\cdots$.

\begin{figure}[t]
\centering
\includegraphics[width=\columnwidth]{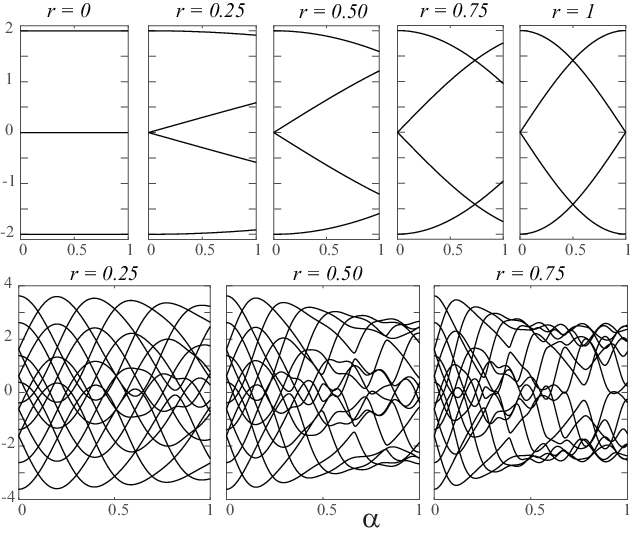}
\caption{(Color online) The origin of the butterfly shape is shown with the gauge dependence varying from $f(n)=-n^r$, with $r\in[0,1]$. Upper panels show for $2\times 2$ and lower for $4\times 4$ cylindrical lattice.}
\label{Fig4-exp}
\end{figure}

\begin{figure*}[t]
\centering
\includegraphics[width=\textwidth]{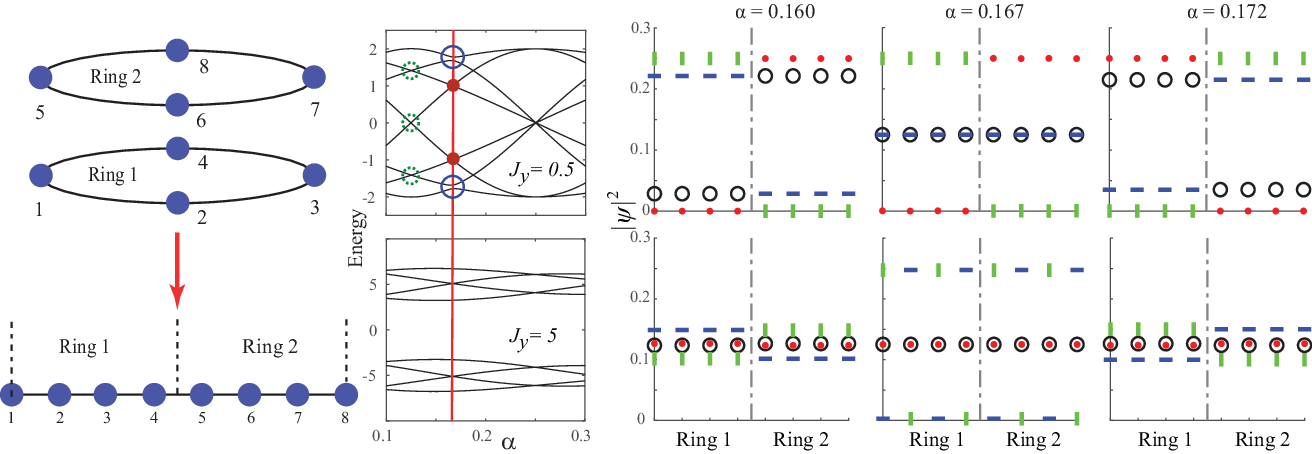}
\caption{(Color online) The schematic uses a $2\times 4$ cylindrical lattice with $f(n)=-n, g(m)=0$ to illustrate how we plot the eigenstates, by mapping the 2D lattice sites to a 1D sequence. The spectra shown at $J_x=1$ for two separate inter-ring coupling $J_y$  trace avoided crossings (solid blue circles), intersections that slide by (dotted green circles) and persistent degeneracies (solid red dots, also tracked with vertical red line). The lowest four eigenstates are shown at $\alpha=1/6$ and its neighborhood, where avoided crossings and persistent degeneracies are both present. In this and other plots we will use the legend having \emph{increasing} energy (\emph{black circles, blue horizontal bar, green vertical bar, and red dots}). At low coupling $J_y=0.5$ (top row) the avoided crossings mark blending of the corresponding states in initially isolated rings, while at persistent degeneracies states remain localized in individual rings. Higher $J_y$ (bottom row) forces stronger blending of the states. At the degeneracy point itself, oscillations (lower middle panel) indicate a linear combination of degenerate states.}
\label{Fig5-2ring_states}
\end{figure*}

Now, we turn to the critical case $J_x=J_y$, but first consider a constant hopping phase, $f(n)=-1, g(m)=0$ or $f(n)=0,g(m)=1$. For the box boundary as well as for a cylinder with the $y$ gauge, shown in the first two rows of Fig.~\ref{Fig3-Spectralcombo}, even with a phase along the axial direction there is no $\alpha$ dependence and the degeneracies are only partially lifted by the added coupling. This shows that bidirectional coupling and the $\alpha$ dependent phase are trumped by the box-boundary conditions. However, when we use $f(n)=-1, g(m)=0$, we see in the third row of Fig.~\ref{Fig3-Spectralcombo} that the degeneracy is completely lifted, the $\alpha$ dependence reflects the sinusoidal pattern for the 1D ring in Eq.~\ref{1Dring}  with the degenerate lines simply spreading apart uniformly. Notably, the butterfly pattern is still absent. The dependence of the phase on the transverse lattice index, associated with the curl is essential to creating the signature butterfly shape. This is borne out by the fact that it is present in the lattice with box boundary conditions in both directions. The phase on $J_x$ needs to depend on the $y$ coordinate and vice-versa. Thus, when $f(n)=-1, g(m)=0$ that shape is absent, but appears when $f(n)=-n, g(m)=0$.  The origin of this characteristic shape can be demonstrated by varying between these limits setting $f(n)=-n^r$ with $r\in [0,1]$ (Note $r$ cannot be a multiplicative factor, in which case it would simply rescale $\alpha$). For the simplest case of a $2\times 2$ cylinder, the four eigenstates are given by
\begin{equation}
E = \pm\sqrt {J_x^2 + J_y^2 \pm 2 J_x J_y \cos[(-1+2^{r}) \alpha \pi]}
\end{equation}
Plotting them versus $\alpha$ for $r=0,1/4,2/4,3/4,1$ we see the gradual morphing of the spectrum from having no dependence on $\alpha$ to the emergence of a skeletal butterfly shape in the upper row of Fig.~\ref{Fig4-exp}. The expression shows that the spectral lines have a period of $2/(-1+2^r)$, which ranges from no period for $r=0$ to a period of $2$ for $r=1$ that corresponds to the usual butterfly shape. Thus over the range $\alpha\in[0,1]$, the spectrum undergoes a half period of oscillation and the central cinch of the butterfly corresponds to where the \emph{cosine} has a zero. With more sites, in the $r=1$ limit we have multitude of terms with $\cos(k\alpha \pi)$ with natural numbers $k$, having a common extremum at the midpoint, resulting in the emergence of the central cinch as illustrated in the lower row of Fig.~\ref{Fig4-exp}.

\section{Spectral features and effects on Eigenstates}

We now identify distinctive spectral features and their impact on the eigenstates.  We represent the eigenstates by lining up the rings that comprise a cylindrical lattice side by side according to $M\times(n-1)+m$ with ring index $n\in{1\cdots N}$ and site index on each ring $m\in{1\cdots M}$.  This is illustrated in Fig.~\ref{Fig5-2ring_states}(a).  We will assume Landau gauges, but in a cyldinder $x$ and $y$ gauges lead to different features, which we discuss in turn.

\subsection{Landau $x$ gauge}

We consider the Landau $x$ gauge first, also assumed in Fig.~\ref{Fig5-2ring_states}. Our analysis is based on the origin and evolution of intersections of the various spectral lines. This is particularly convenient for this gauge, because in the limit $J_y=0$ of decoupled rings, the spectrum still depends on the field variable $\alpha$ due to the periodic boundary condition as seen on the left column of the last row of Fig.~\ref{Fig3-Spectralcombo}. In that limit it is possible to uniquely associate each spectral line with a specific state $m\in 1,2,\cdots M$ of a specific ring $n\in 1,2,\cdots N$, as captured by the expression in Eq.~(\ref{1Dring}). Starting from that limit, as $J_y$ is increased coupling the rings to create a true cylinder, we trace the evolution of those intersections and the associated eigenstates.

We observe three primary types of behavior based on the type of intersections in the limit of $J_y=0$ when the rings are decoupled: (I) Same state ($m$) from different rings ($n$): Gaps open up creating avoided crossings as $J_y$ is turned on and increased; (II)  Different states ($m$) from the same ring ($n$): These intersections slide past each other, intersecting at different values of $\alpha$ as $J_y$ changes and (III) Different states ($m$) from different rings ($n$): These intersections persist at exactly the same value of $\alpha$ for any value of $J_y/J_x$. All three cases are illustrated in Fig.~\ref{Fig5-2ring_states}.  In the limit of $J_y\gg J_x$ the states approach the limit of Eq.~(\ref{1Dstrip}) the spectral lines rearrange into $M$ bands, with each band having a spectral `line' from each of the $N$ values at $\alpha=0$ when $J_y=0$.  Thus, the intersecting lines in cases I and II end up in different bands but those in case III end up in the same band.

\emph{Persistent Degeneracies}:  These are degeneracies that persist at the same value of $\alpha$, insensitive to the coupling constants, that is for any ratio $J_y/J_x$. These are marked by red dots in Fig.~\ref{Fig5-2ring_states}.  Since gaps \emph{never} open up at any point of intersection between these spectral lines, these pairs of lines remain braided in exactly the same way even though their individual shapes change drastically, as seen in Fig.~\ref{Fig5-2ring_states}. In the opposite limit, $J_y\gg J_x$ or equivalently $J_x\rightarrow 0$, such pairs inevitably end up in the same band, but with the lines having new identities as arising from the same state but from different weakly coupled strips. We have shown elsewhere \cite{Das_Brooks_short} that these occur at $\alpha$ that satisfy $\alpha M (N+1)=k$ with natural number $k$.  For $J_y\rightarrow 0$ the state associated with each line retains the same form across the intersection, for example in Fig.~\ref{Fig5-2ring_states} the upward slanting spectral line marks the state localized on ring 1 for $\alpha$ both below and above the intersection, although the markers swap based on the energy ordering. Notably the same behavior continues even at very different coupling ratios.

\begin{figure}[t]
\centering
\includegraphics[width=\columnwidth]{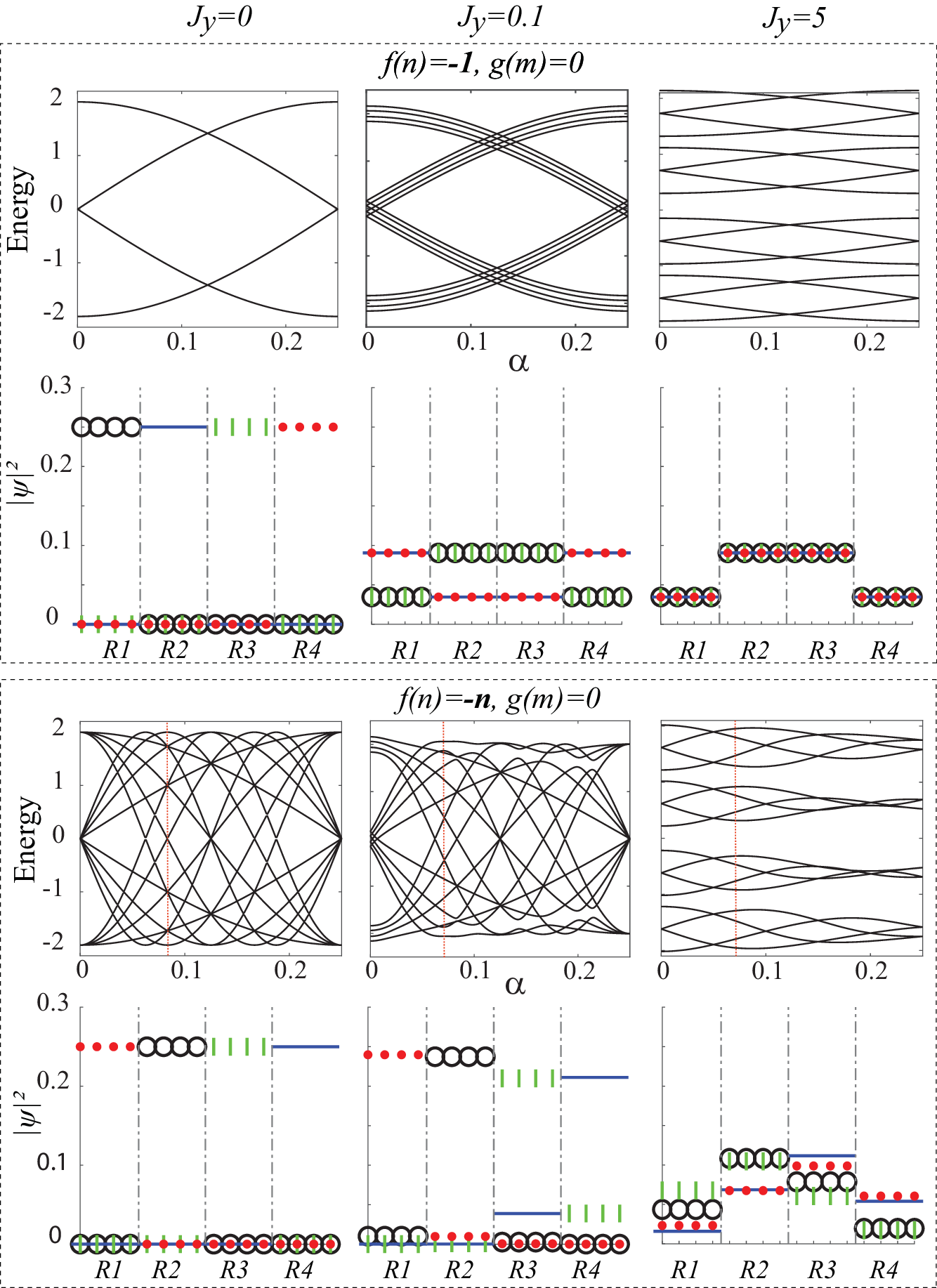}
\caption{(Color online) The effect of the phase variation among the rings is shown by contrasting the case of $f(n)=-1$ (top rows) with $f(n)=-n$ bottom rows. (\emph{Row 1}) As the inter-ring coupling is increased from the left to right, for $f(n)=-1$, the degeneracies are lifted independent of $\alpha$, and the spectral lines separate uniformly. (\emph{Row 3}) For $f(n)=-n$ the relative phase between the rings causes strong dependence on $\alpha$, causing gaps to open up at specific values leading to avoided crossings. (\emph{Row 2}) In the absence of phase difference between rings, even a small coupling $J_y$ causes the eigenstates to mix; (\emph{Row 4}) the phase difference resists mixing, which occurs selectively around avoided crossings only, shown here for one at $\alpha=0.0715$; mixing is more pronounced at stronger $J_y$.}
\label{Fig6-cylvscone}
\end{figure}

\emph{Avoided Crossings}: As the inter-ring coupling $J_y$ is switched on and increased, gaps open up at erstwhile intersections between spectral lines for the corresponding states (same $m$) from different rings (different $n$). These take the form of avoid crossings marked by blue circles in Fig.~\ref{Fig5-2ring_states}.  The eigenstates confirm the characteristic feature of avoided crossings, a blending of states that were isolated in different rings. This is analogous to what happens when gaps open up in a periodic lattice as the coupling between the sites is strengthened; here each ring behaves like a site in the axial orientation of the cylinder. At low coupling $J_y=0.5$ in the figure, there is complete blending of the states at the $\alpha$ value of closest approach, while away from it the states become more localized. At high inter-ring coupling $J_y=5$, the gaps widen to eventually create different bands.

In order to understand these avoided crossings better, we compare them with the case where the phase factor is the same for all the rings $f(n)=-1$, in Fig.~\ref{Fig6-cylvscone}.  For this case, when $J_y=0$, the spectral lines of each state is degenerate for all the rings for any $\alpha$. As soon as we have any coupling between the rings $J_y>0$, the degenerate lines separate uniformly, and the gaps have no dependence on $\alpha$, even though the spectral lines themselves clearly vary with it. This can be contrasted with the case $f(n)=-n$, where the degeneracies occur only at certain values of $\alpha$ and gaps open up prominently at those values leading to avoided crossings. The behavior of the eigenstates underscores this difference. For $f(n)=-1$ with identical phase variation in the rings, even a very small inter-ring coupling is enough to instantly blend all the states at any value of $\alpha$.  This is contrasted with the case of $f(n)=-n$ where such blending occurs gradually and is most pronounced at $\alpha$ values where the avoided crossings occur.

Thus, the phase difference between the rings presents resistance to the blending of corresponding eigenstates across the rings, and the ring dependence of the phase is essential for avoided crossings to occur. Implicit here is a hierarchy of avoided crossings based upon the separation of the rings. Larger coupling $J_y$ is required to open up significant gaps at intersections of spectral lines associated with initially decoupled rings that are farther apart.  Furthermore with increasing $J_y/J_x$  the blending between states occurs at $\alpha$ values farther from the avoided crossing, eventually merging the same states in different rings to transition to the limit of decoupled strips.

\begin{figure}[t]
\centering
\includegraphics[width=\columnwidth]{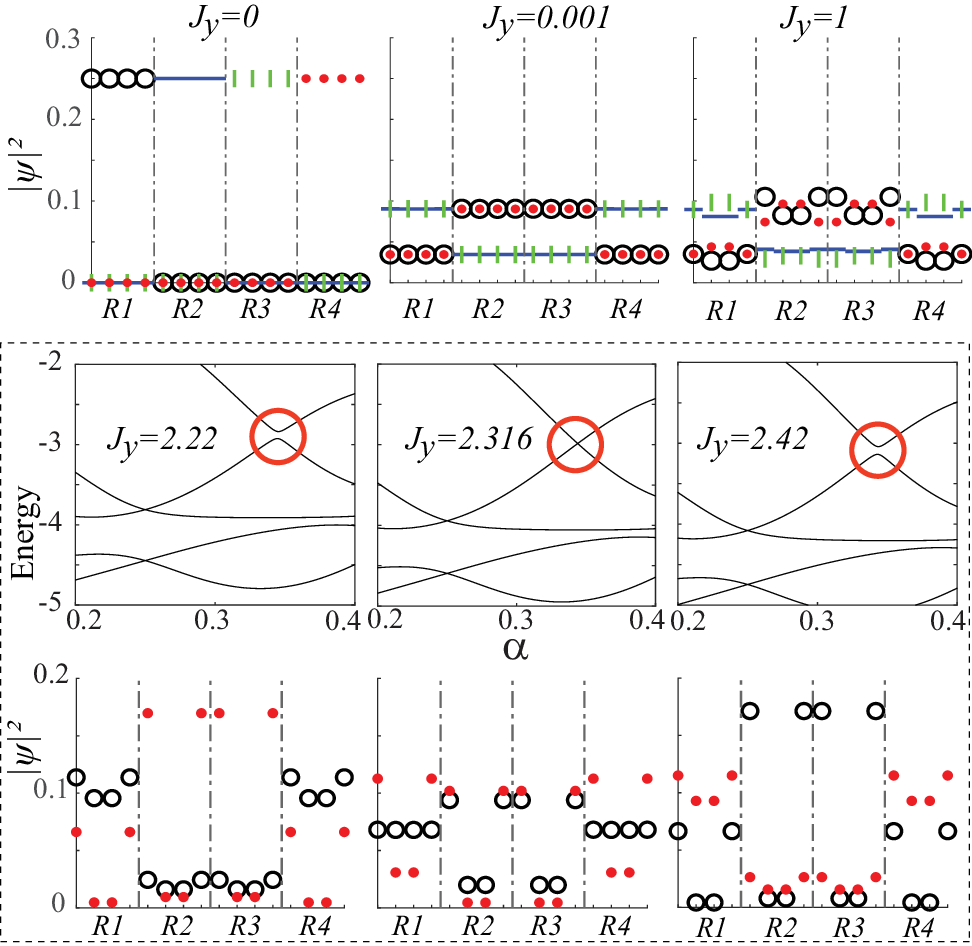}
\caption{(Color online) (\emph{Row~1}) For $y$ gauge there is mixing of states among rings even at weak coupling $J_y$ between rings for fixed $J_x=1$. The left two panels are identical for both $f(n)=0, g(m)=1$ and $f(n)=0, g(m)=m$; differences emerge at stronger coupling in the right panel,  the phase variation $g(m)=m$ creates modulation within each ring. (\emph{Row~2}) Degeneracies can appear in the spectrum  transiently as the coupling ratio $J_y/J_x$ is varied, here at $\alpha=0.345$, then parting to create gaps that resemble avoided crossings. (\emph{Row~3}) Across that degeneracy point there is a switching of the states corresponding to the energies above and below the gap.}
\label{Fig7-yperiodic}
\end{figure}

\subsection{Landau $y$ gauge}
When the gauge is along the open direction, dependence on $\alpha$ requires nonvanishing coupling in both $J_x$ and $J_y$ as seen from Fig.~\ref{Fig3-Spectralcombo}.  Starting from the limit of isolated rings $J_y=0$, as soon as coupling $J_y=0.001$ is introduced, even for such small inter-ring coupling, there is immediate blending of the eigenstates, regardless of whether $g(m)=1$ or $g(m)=m$, meaning that this occurs whether or not the phase factor associated with the $y$ gauge varies from one open strip to the next. This can be seen in the upper row of Fig.~\ref{Fig7-yperiodic}.  However, as the coupling between rings is increased, to $J_y=1$ in the figure, the presence of phase variation $g(m)=m$ among the strips associated with the $y$ gauge creates modulation of the eigenstates within each ring. This coincides with $\alpha$ dependence in the spectrum becoming more pronounced.

\emph{Faux Avoided Crossings}: In the $y$ gauge, we see spectral features that appear to be avoided crossings, with multiple examples in Fig.~\ref{Fig3-Spectralcombo}. However, in both limits $J_y=0, J_x\neq 0$ and $J_y\neq0, J_x =0$, the dependence on $\alpha$ is lost, so there are no `crossings' to start with unlike in the $x$ gauge.  Thus, true avoided crossings do not occur in the $y$ gauge, as supported by the behavior of the states at gaps that resemble avoided crossings of two lines. The eigenstates at and around those gaps show complete blending even when one of the couplings is extremely small, as can seen in the upper row of Fig.~\ref{Fig7-yperiodic}; this clearly does not depend on $\alpha$ in that limit. This is very different from the behavior at true avoided crossings where the blending gets less pronounced away from the point of closest approach of the relevant spectral lines as we saw in Fig.~\ref{Fig5-2ring_states}. Rather the behavior is analogous to that for the $f(n)=-1, g(m)=0$ limit of the $x$ gauge when all the rings have the same phase: In the limit of either coupling vanishing, the relevant spectral lines are degenerate with  identical (and here, no) dependence on $\alpha$, so that even small coupling that lifts the degeneracy will blend the states uniformly across all values of $\alpha$.

\emph{Coupling dependent Degeneracies}: We observe another phenomenon with $y$ gauge  not apparent with the $x$ gauge: degeneracies that occur at specific values of the coupling constants $J_y/J_x$ manifest as transient intersections of pairs of spectral lines at a certain values of $\alpha$ as the coupling ratio is varied. One such  degeneracy shown in Fig.~\ref{Fig7-yperiodic}, occurs at $\alpha=0.345$ and we see that the degeneracy lifts above and below the specific $J_y=2.316$ keeping $J_x=1$ fixed.  In the lowest row we plot the states above and below the gap for the $\alpha$ with the degeneracy, as we vary the coupling across the degeneracy point. There is a switching of the states associated with the energy above and below the gap, but with no fundamental change in behavior in the relevant states. This behavior occurs also with box boundary conditions for both gauges, hence is not due to periodic boundary condition. Furthermore, at fixed coupling ratio such at $J_x=J_y=1$ plotted in Fig.~\ref{Fig3-Spectralcombo}, some of the `faux' avoided crossings can become transiently degenerate as the coupling ratio is varied.

\begin{figure}[t]
\centering
\includegraphics[width=\columnwidth]{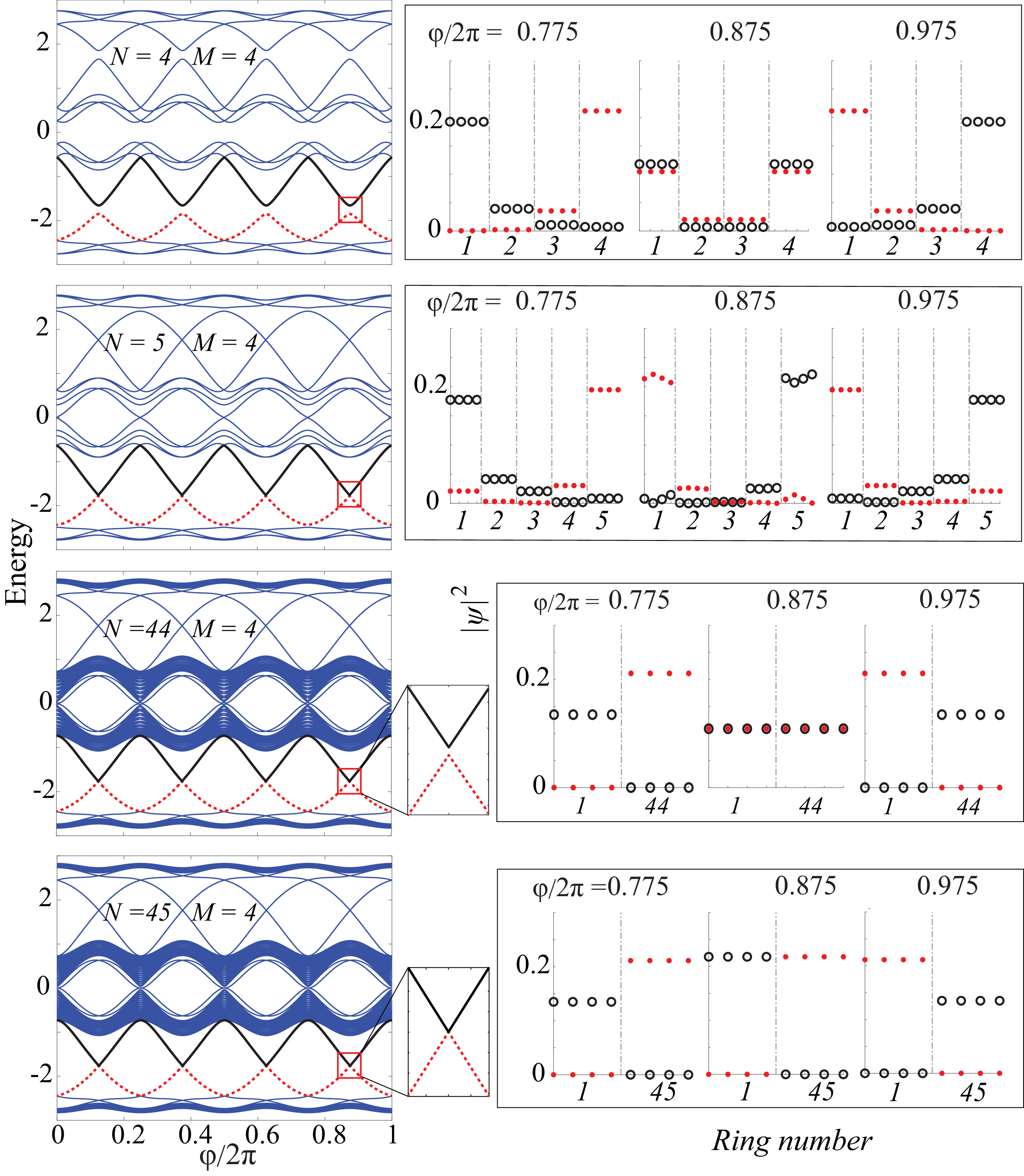}
\caption{(Color online) The spectrum of a cylindrical lattice is plotted as a function of an added axial field parameterized by $\varphi$ shown for the critical case $J_x=J_y=1$ and $\alpha=1/4$. Sites per ring $M=4$ is commensurate with $\alpha$ while number of rings $N$ is variable. With a few rings, when $N$ is commensurate with $\alpha$, visible avoided crossings appear in the intersections of spectral lines (thick black line and red dotted) that span the band gaps; $N=4$ and $N=5$ are contrasted. That behavior persists at large ring numbers, illustrated with $N=44$ and $45$, but with smaller width seen only on zooming in. On the right, the eigenstates (black open circles and red dots,  representing the lower and higher energy states respectively) that correspond to those spectral lines are plotted for a set of $\varphi$ centered about such an intersection marked by a box in the spectrum plots. They are localized on the edge of the cylinder; for the larger system in the bottom rows, only the edge rings are shown. Avoided crossings  lead to a blending of edge states at the point of closest approach as seen in \emph{rows 1} and \emph{3}; which does not happen otherwise in \emph{rows 2} and \emph{4}.}
\label{Fig8-edge_states_gap}
\end{figure}

\section{Edge states of Cylindrical Lattices}

We now examine the features of edge states that have played a significant role in the understanding of the quantum Hall effect \cite{Laughlin-PRB-QHE, Halperin-QHE}. Properties of edge states have been studied exhaustively in the context of macroscopic conductors,  necessarily cylinders with large number of azimuthal sites. Even with ultracold atoms, studies have focussed on large cylinders \cite{Lacki-cylinder}. Here, we discuss certain behavior that emerge in cylindrical lattices of a few sites, in the context of Figs.~\ref{Fig8-edge_states_gap} and \ref{Fig9-top_ins_tansition}.

\subsection{Gaps in the Edge spectrum}

In this context, it is better to represent the spectrum as a function of the phase factor along the closed orientation of the cylinder, $J_x\rightarrow J_x e^{i\varphi}$ that corresponds to an extra axial field, the essential feature of the Laughlin model for the quantum Hall effect \cite{Laughlin-PRB-QHE}. We will refer to that axial magnetic field to include both real or synthetic kind. In the spectrum plotted in Fig.~\ref{Fig8-edge_states_gap}, as a function of this axial field, $\varphi$ the edge states appear as spectral lines that traverse the gap between bulk bands; that is the feature that gives them their essential role in quantum Hall physics. Plots of the corresponding eigenstates show that those states are indeed mostly localized on the edge rings of the cylinder particularly at the $\varphi$ values near the intersections of the edge state spectral curves.

With a few azimuthal sites $M=4$, we notice in Fig.~\ref{Fig8-edge_states_gap} that the intersections of the edge state curves can become avoided crossings. We find that this occurs when N and $\alpha$ are commensurate, as in $N\alpha=k\in\{1,2,\cdots\}$. Thus for $\alpha=1/4$ used in the figure, avoided crossings appear for $N=4$ but not for $N=5$; and they persist even with larger number of rings, appearing for $N=44$ but not $N=45$. This sensitivity to the number of axial sites is among the features that distinguish this from  other gaps in the spectrum reported in some recent works on Hall ladder configurations \cite{Shin_bandgap-closing-PRL,Santoro-synthetic-Hall,Zhou_PRX_Hall_cylinder}. However, when all other parameters are fixed, the width of those avoided crossings  progressively shrinks with increasing number of rings, although still present as seen by zooming in at the crossings. These gaps will therefore be physically relevant only for a small number of rings.

There are other restricting considerations for these avoided crossings to appear and be relevant. Fig.~\ref{Fig8-edge_states_gap} considers $M \alpha = 1$. This choice generally assures that there are only two spectral lines in the lowest band gap for any $\varphi$. Additionally, as implicit in Fig.~\ref{Fig3-Spectralcombo} and related discussion in Sec.~4, the band structure changes as the ratio $J_y/J_x$ varies. At low values, we have weakly coupled rings and there are $M$ bands, whereas at high values, we have a set of weakly coupled strips and there are $N$ bands. This means a meaningful interpretation as a cylinder with edge states requires a low $J_y/J_x$. When $M \alpha = 1$ the edge states in the lowest gap correspond to the  $N^{th}$ and $(N+1)^{th}$ states. Thus when $N$ and $\alpha$ are commensurate, the band gaps appear between those states creating the avoided crossings in the edge states.

\begin{figure}[t]
\centering
\includegraphics[width=\columnwidth]{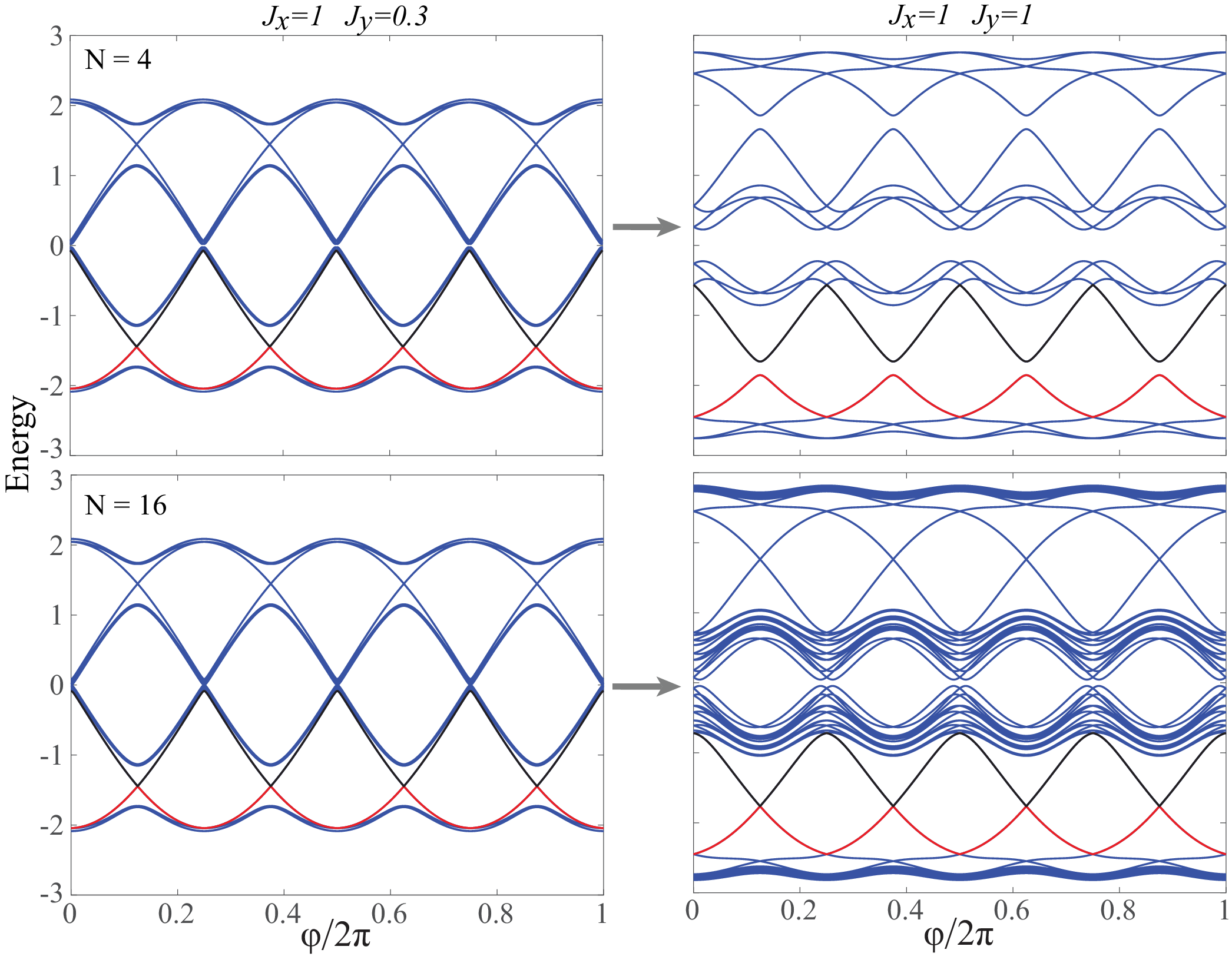}
\caption{(Color online) The avoided crossings that occur in the edge state spectral lines widen by increasing the inter-ring coupling $J_y$, but the effect diminishes significantly for a taller cylinder with larger number of rings $N$; comparison is made between $N=4$ and $N=16$.}
\label{Fig9-top_ins_tansition}
\end{figure}

The width of the avoided crossings can be widened by increasing $J_y/J_x$. But for larger number of rings, that needs to be progressively larger. This is illustrated in Fig.~\ref{Fig9-top_ins_tansition} where, keeping all else fixed, increasing the number of rings from $N=4$ to $N=16$, suppresses the avoided crossing, although it still exists if one zooms in since $N\alpha$ remains a positive integer with $\alpha=1/4$. However, making $J_y/J_x$ larger places the system in the limit of weakly coupled vertical strips rather than a cylinder, and the edge states carry less meaning. If $\alpha$ is reduced, creating the avoided crossings requires proportionately larger number of rings $N$ and hence the width of the avoided crossing will shrink, requiring large $J_y/J_x$. This means larger magnetic fields are preferable for the observation of avoided crossings.

Keeping $\alpha$ fixed, if we increase the number of sites in the ring, such that $M \alpha >1$ and is commensurate, then more spectral lines appear in the band gap with more frequent intersections and even multiple intersections occurring at the same $\varphi$ value. This complicates the identity of the edge states, but the avoided crossings still appear only on specific intersections of the spectral lines in the band gap, but now at lines numbered $NM\alpha$ and $NM\alpha+1$. The size of the width of the avoided crossing remains unaffected by the number of azimuthal sites.

\subsection{Effect on the states}

The appearance of the gaps implies a change in behavior of the system. Edge states which traverse the space between bulk bands is a characteristic of the integer quantum Hall effect, an example of topological insulators.  When avoided crossing open up between these states then the band gap always remains open, and this becomes a trivial insulator. Thus the emergence of the avoided crossing can be a signature of a change in the topological properties of the system \cite{A5}. In systems with a large number of lattice sites, where the thermodynamic limit can be assumed, change of topological properties can be captured by a change of the quantized Chern numbers for the bands involved. However, for small finite periodic lattices that we study here, Chern numbers cannot be computed accurately and are not necessarily quantized \cite{Suzuki-Chern,Chern-1}. Generally, sharp phase transitions are characteristic of thermodynamic limit, and transitions can be gradual in finite systems  \cite{Das-PRL-localization}.

The avoided crossings that appear in the spectral lines spanning the gaps have interesting implications on the corresponding edge states. In Fig.~\ref{Fig8-edge_states_gap} for each spectrum shown, we plot the states corresponding to the spectral lines that traverse the gaps, around one of the marked intersections/avoided crossings. In the case of large number of rings, $N=44,45$, only the edge rings are shown. Regardless of the number of rings, we can see that the behavior of the spectrum directly influences the states. These states are localized on the edges.

The two relevant spectral lines are differentiated by the energy, lower (dotted red) and higher (thick black), regardless of the presence or absence of a gap. The corresponding eigenstates are marked with red dots and black circles respectively. What stands out is that the state localized on a particular edge follows a specific edge state line across the intersection, and even when it becomes an avoided crossing. This is underscored by the switching of red dots $\rightarrow$ black circles in Fig.~\ref{Fig8-edge_states_gap} between the edge rings due to the reordering of the energy across the intersection or avoided crossings. Similar behavior for avoided crossing is seen in Fig.~\ref{Fig5-2ring_states}.

The appearance of the avoided crossing in the edge state spectral lines has another interesting effect. Adiabatic time evolution along the lower energy (red dotted) line, transfers the density localized on one edge to the other; and likewise, if one followed the upper energy (thick black line). This seems rather like  Laughlin's argument for the charge transfer between the edges due to a variable flux through the cylinder,  except of course, that this is the situation where the system would be a trivial insulator. That is because there is no net transfer when averaging over all the states in the lowest band. However, with ultracold atoms in cylindrical lattices, one could engineer a system where all the atoms are on the edge state of the lower band, and then a variation of the flux can adiabatically pump the atoms from one edge to the other.

\begin{figure}[t]
\centering
\includegraphics[width=\columnwidth]{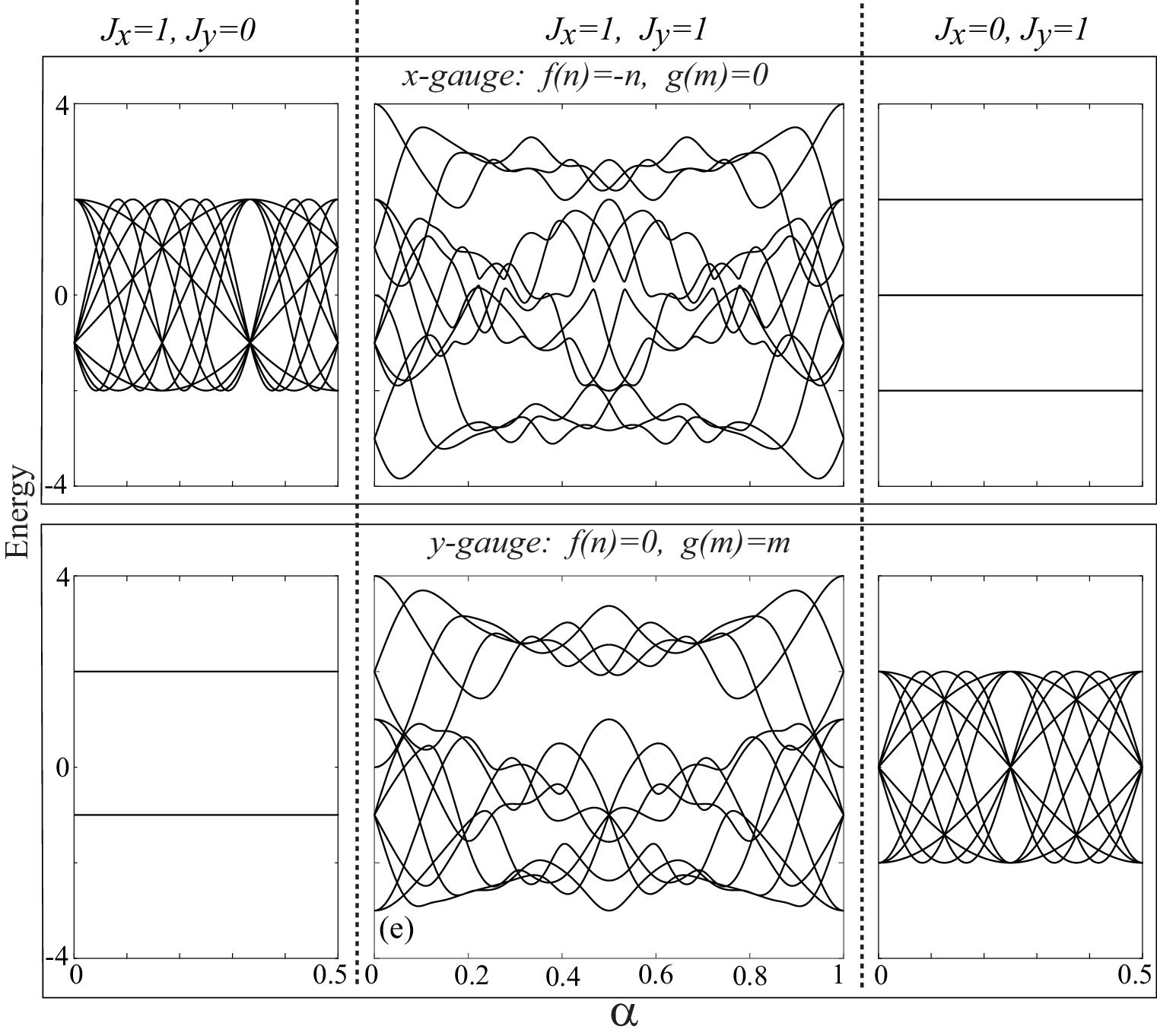}
\caption{(Color online)  The plots are a counterpart of Fig.~\ref{Fig3-Spectralcombo} for a torus that compares Landau $x$ and $y$ gauges, but with asymmetric site number $N\times M=4\times 3$. Qualitatively, the critical case of $J_x=J_y=1$ in the central column are similar to that for a cylinder. Due to periodic boundary conditions in both orientations, the two gauges display similar behavior; particularly in the limiting cases, where one of the coupling vanishes, the spectrum shows dependence on $\alpha$ when the phase associated with the gauge is along the coupled direction, and the eigenvalues are given by Eq.~(\ref{1Dring}).}
\label{Fig10-torus_spectrum}
\end{figure}

\section{A Toroidal Lattice}

We will now apply and extend our results for a cylindrical lattice  to a lattice with a torus configuration. This can be done by connecting the open ends of the cylinder. We disregard distortions or stretching that could occur in a real torus in doing so, and model it simply as having periodic boundary condition in both $x$ and $y$ directions. Most of the considerations for the cylinder have counterparts for a torus.  Therefore, we will only discuss the principal features and focus on the differences.

\emph{Spectrum}: In analogy to  Fig.~\ref{Fig3-Spectralcombo}, we examine the features of the  energy spectrum for a torus. As for the cylinder we vary the gauge between Landau $x$ and $y$ gauges as well as the coupling strength in the two directions, and the results are plotted in Fig.~\ref{Fig10-torus_spectrum}. With periodic boundary condition in both directions, different number of sites along the two axes, $N=4$ and $M=3$, are used to discern differential behavior.

The characteristic butterfly pattern is present for the torus as well as seen in panels (b) and (e), when the coupling constants are the same or comparable in both orientations, and the same explanation applies as for the cylindrical configuration discussed earlier in the context of Fig.~\ref{Fig4-exp}. Due to the periodic boundary conditions in both directions, the butterfly pattern is similar for both Landau gauges, the differences are mostly due to the different number of lattice sites, $N\neq M$.

Other differences appear in the limit of relatively high coupling in either orientation.  In the Landau $x$ gauge, in the limit of $J_x=0,J_y=1$ there are less energy bands, with three instead of four in panel (c), due to additional degeneracies. For the Landau $x$ gauge, in that limit, there was no dependence on the field $\alpha$ for the cylinder, but with a periodic boundary condition in the $y$-direction as well, there is dependence on $\alpha$ in that limit as can be seen in panel (f).

\emph{Persistent Degeneracies}:  Unlike for cylinders, persistent degeneracies appear for both $x$ and $y$ gauges underscoring the essential role of periodic boundary condition present in both directions now \cite{Das_Brooks_short}, but with an interesting distinction between the gauges.  For $x$ gauge, if we convert a cylinder to a torus by introducing a periodic boundary condition in the open orientation, the persistent degeneracies continue to reside at the same values of $\alpha$ even though the spectrum itself changes. For the $y$ gauge, new persistent degeneracies are emergent at values  of $\alpha$ set by the same expression as for the $x$ gauge, but with $M \leftrightarrow N$ switched, $\alpha N (M+1)=k$.  This is tied to the emergence of $\alpha$ dependence in the $J_x=0, J_y=1$ limit in Fig.~\ref{Fig10-torus_spectrum}. Therefore, for persistent degeneracies to appear, the gauge needs to be along a direction with periodic boundary condition.  A cylinder with $y$ gauge display cinches tied to the butterfly shape as addressed in Fig.~\ref{Fig4-exp}, and in the spectrum for its torus counterpart, they continue to define the envelope as new intersections emerge including those corresponding to persistent degeneracies.

\begin{figure}[t]
\centering
\vspace{0.075\linewidth}
\includegraphics[width=\columnwidth]{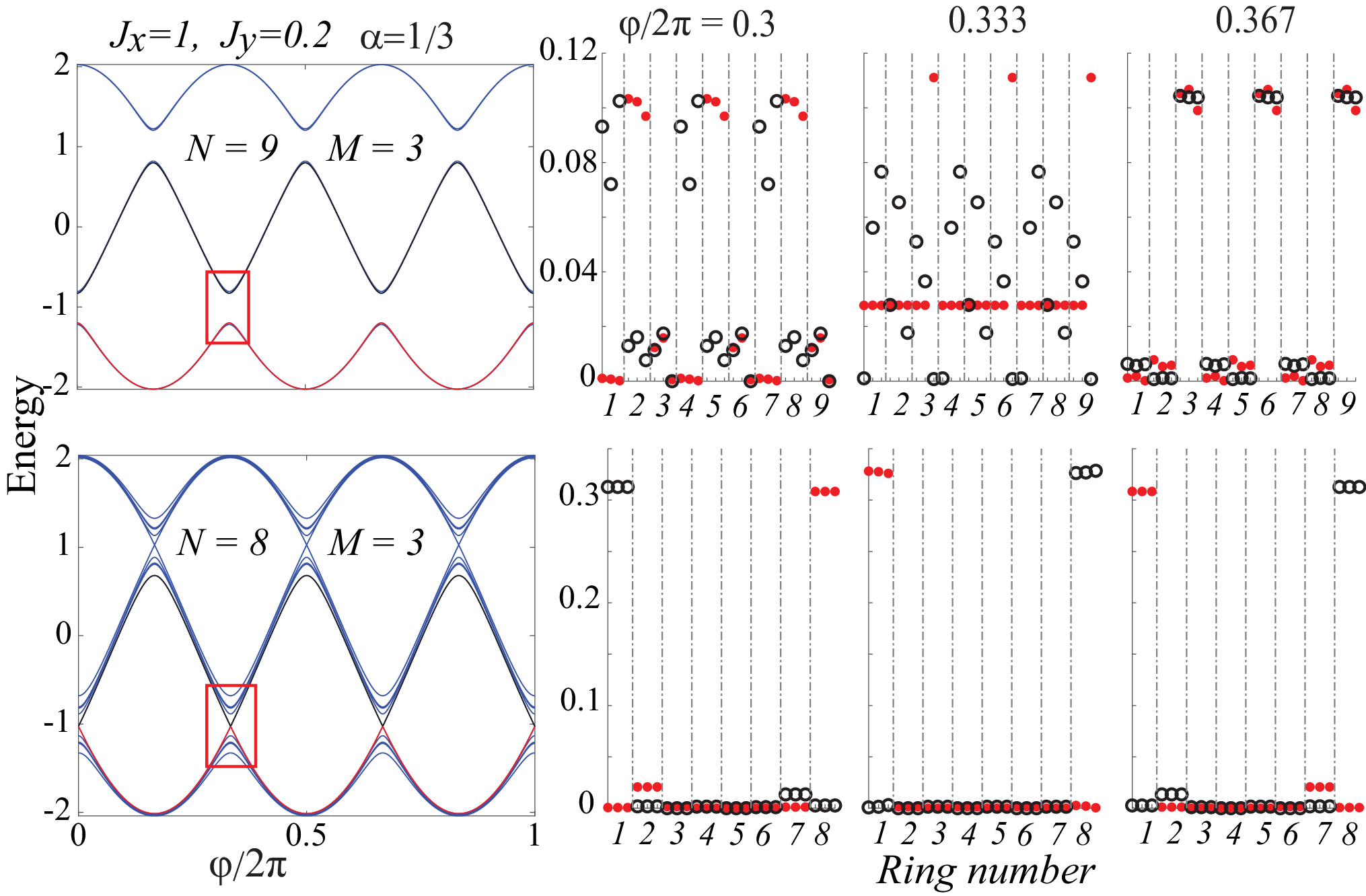}
\caption{(Color online) With the same convention used for cylinders, assuming $x$ gauge, the spectrum is plotted versus an additional phase along the axis in the $x$ direction. When site number $N$ in the $y$ direction without the gauge factor, is commensurate with $\alpha$, there are no continuous inter-band spectral lines (upper row) they exist only when incommensurate, intersecting pairs of such lines appear (lower row).  This is similar to the behavior seen for edge states in cylinder in Fig.~\ref{Fig8-edge_states_gap}. States corresponding to these `edge' spectral lines are plotted for some of the phase values around where such intersections happen, using the same format as for cylinders noting that first and last rings are adjacent and coupled in a torus. Commensurate/incommensurate cases correspond to delocalized/localized behavior of the states.}
\label{Fig12_torus_edge_states}
\end{figure}

\emph{Torus `edge' states}: A torus by construction has no edge, so it may seem meaningless to talk of `edge' states. But we find that there are certain parallels in the spectrum and the associated states of a torus that are the counterparts of the corresponding cylindrical lattice.  For a torus, with periodic boundary conditions in both directions, there is no qualitative difference between the two Landau gauges. We label $N$ and $M$ analogously to the $x$ gauge for a cylinder, so that $M$ is the number of sites along the direction of the gauge, and we will refer to $N$ as the number of rings for easy comparison although now  rings occur in both orientations.

We illustrate our main findings in Fig.~\ref{Fig12_torus_edge_states} where we plot the spectrum as a function of an added field along the axis in the $x$ direction creating an added phase $J_x\rightarrow J_xe^{i\varphi}$ in analogy with  Fig.~\ref{Fig8-edge_states_gap} for the cylinder. As we did for the cylinder in Fig.~\ref{Fig9-top_ins_tansition}, we increase the coupling ratio $J_y/J_x$ but stay at relatively low values. Intersections present when $J_y/J_x=0$ lead to avoided crossings, but of comparable widths at noticeably smaller values than for a cylinder; hence we consider a smaller value $J_y/J_x=0.2$ for the torus. We use $M=3$ and $\alpha=1/3$ so that they are commensurate, $M\alpha=1$, and then edge states in a cylindrical latttice would correspond to states $N$ and $N+1$. We can see that as with the cylinder when $N\alpha$ is commensurate with $N=9$, these `edge' states present avoided crossings, whereas when $N\alpha$ is incommensurate with $N=8$, there is no such gap and the two relevant spectral lines intersect and traverse the band gap.

In the case of a torus, considering that there are no actual edges, we look for other implications of this difference in behavior by plotting the states. We keep the same format as for the cylinder, we described in Fig.~\ref{Fig5-2ring_states}, except that we need to keep in mind that the first ring and the last ring are adjacent. We find an interesting difference in behavior between the commensurate and the incommensurate cases described above: When $N\alpha$ is incommensurate the states are strongly localized in that direction, as seen in the lower row of Fig.~\ref{Fig12_torus_edge_states}. With $N\alpha$ commensurate, as shown in the upper row of that figure, the states $N$ and the $N+1$  are delocalized and show certain periodic behavior in the direction indexed by $n$, particularly in the neighborhood of the avoided crossing. The delocalization is actually present in all the states, pronounced by the near degeneracy of the states in each band. This delocalization and periodicity can be understood as a manifestation of lattice periodicity matching the magnetic translation period \cite{Zak-1}.


\section{Outlook and Conclusions}

We conducted a comparative study of the spectrum and states of small 2D lattices having cylindrical and toroidal topologies  subject to an effective gauge field orthongonal to the lattice. Our considerations although more generally applicable, are motivated by synthetic gauge fields in ultracold atoms, where such configurations are now being actively examined in experiments. For cylindrical lattices, by considering the commonly used Landau gauge, we demonstrate how the orientation of the gauge relative to the periodic boundary condition fundamentally alters the behavior of the system.  The equivalence is restored at the values of the field parameter $\alpha$ that are commensurate with the number of lattice sites in the periodic direction. By comparing the spectra with and without periodic boundary conditions and for limiting cases of vanishing inter-site coupling  in one or the other orientation, we could isolate the impact of various competing factors that lead to the well-known butterfly like spectrum.

Focusing on small lattices enabled classifications of the various degeneracies in the spectra manifest as intersections of spectral lines plotted as functions of the field parameter $\alpha$ or an additional axial field parameterized by $\varphi$. Key among these are degeneracies that evolve into avoided crossings, as well as persistent degeneracies that remain invariant in several ways with respect to changes in the coupling constants of the lattice. We discuss the behavior of the associated eigenstates, which in turn show that the very presence of these features depend on the synthetic gauge chosen.

For small size cylindrical lattices, we could identify the spectral lines that span the band gaps as states that are indeed localized on the edges, and hence their nomenclature as edge states. These have been shown to play a crucial role in the quantum Hall effect \cite{Halperin-QHE,Laughlin-PRB-QHE}.  Here we found that as a function of the added axial field $\varphi$, degenerate points in pairs of such spectral lines can develop avoided crossings when the number of coupled rings that comprise the cylinder is commensurate with the field variable $\alpha$. This in effect means that a band-spanning spectral line that leads to the characteristic features associated with the integer quantum Hall effect \cite{A2} would be lost. Such effects are sensitive to system size and are particularly prominent in small lattices.

A small toroidal lattice displays many of the same qualitative behavior as  a cylindrical one. Specifically, when the gauge is along the periodic direction of a cylinder, closing it to a torus does not alter the number and appearance of persistent degeneracies; however such degeneracies emerge in a torus while absent in a cylinder when the gauge is in the other orientation.  Notably, when we apply the same conditions that for a cylinder caused edge state spectral lines to transition from toplogical to trivial insulating behavior, we find that there is a transition between localization and delocalization of the relevant eigenstates in  a torus.

The results in this paper can provide guidance in analyzing the behavior of cylindrical and toroidal lattice structures with synthetic gauge fields that are gaining interest in experiments with ultracold atoms.  The behavior of the spectrum and the quantum states as a function of various system parameters and factors, directly impact the dynamical behavior of these systems. Examining such time-dependent behavior particularly in the adiabatic regime will be part of our continuing research.

\begin{acknowledgments} We gratefully acknowledge the support of the NSF under Grant No. PHY-2011767  and PHY-2309025. \end{acknowledgments}


\begin{thebibliography}{38}%
\makeatletter
\providecommand \@ifxundefined [1]{%
 \@ifx{#1\undefined}
}%
\providecommand \@ifnum [1]{%
 \ifnum #1\expandafter \@firstoftwo
 \else \expandafter \@secondoftwo
 \fi
}%
\providecommand \@ifx [1]{%
 \ifx #1\expandafter \@firstoftwo
 \else \expandafter \@secondoftwo
 \fi
}%
\providecommand \natexlab [1]{#1}%
\providecommand \enquote  [1]{``#1''}%
\providecommand \bibnamefont  [1]{#1}%
\providecommand \bibfnamefont [1]{#1}%
\providecommand \citenamefont [1]{#1}%
\providecommand \href@noop [0]{\@secondoftwo}%
\providecommand \href [0]{\begingroup \@sanitize@url \@href}%
\providecommand \@href[1]{\@@startlink{#1}\@@href}%
\providecommand \@@href[1]{\endgroup#1\@@endlink}%
\providecommand \@sanitize@url [0]{\catcode `\\12\catcode `\$12\catcode
  `\&12\catcode `\#12\catcode `\^12\catcode `\_12\catcode `\%12\relax}%
\providecommand \@@startlink[1]{}%
\providecommand \@@endlink[0]{}%
\providecommand \url  [0]{\begingroup\@sanitize@url \@url }%
\providecommand \@url [1]{\endgroup\@href {#1}{\urlprefix }}%
\providecommand \urlprefix  [0]{URL }%
\providecommand \Eprint [0]{\href }%
\providecommand \doibase [0]{https://doi.org/}%
\providecommand \selectlanguage [0]{\@gobble}%
\providecommand \bibinfo  [0]{\@secondoftwo}%
\providecommand \bibfield  [0]{\@secondoftwo}%
\providecommand \translation [1]{[#1]}%
\providecommand \BibitemOpen [0]{}%
\providecommand \bibitemStop [0]{}%
\providecommand \bibitemNoStop [0]{.\EOS\space}%
\providecommand \EOS [0]{\spacefactor3000\relax}%
\providecommand \BibitemShut  [1]{\csname bibitem#1\endcsname}%
\let\auto@bib@innerbib\@empty
\bibitem [{\citenamefont {Thouless}\ \emph {et~al.}(1982)\citenamefont
  {Thouless}, \citenamefont {Kohmoto}, \citenamefont {Nightingale},\ and\
  \citenamefont {den Nijs}}]{TKNN}%
  \BibitemOpen
  \bibfield  {author} {\bibinfo {author} {\bibfnamefont {D.~J.}\ \bibnamefont
  {Thouless}}, \bibinfo {author} {\bibfnamefont {M.}~\bibnamefont {Kohmoto}},
  \bibinfo {author} {\bibfnamefont {M.~P.}\ \bibnamefont {Nightingale}},\ and\
  \bibinfo {author} {\bibfnamefont {M.}~\bibnamefont {den Nijs}},\ }\bibfield
  {title} {\bibinfo {title} {Quantized hall conductance in a two-dimensional
  periodic potential},\ }\href {https://doi.org/10.1103/PhysRevLett.49.405}
  {\bibfield  {journal} {\bibinfo  {journal} {Phys. Rev. Lett.}\ }\textbf
  {\bibinfo {volume} {49}},\ \bibinfo {pages} {405} (\bibinfo {year}
  {1982})}\BibitemShut {NoStop}%
\bibitem [{\citenamefont {Laughlin}(1981)}]{Laughlin-PRB-QHE}%
  \BibitemOpen
  \bibfield  {author} {\bibinfo {author} {\bibfnamefont {R.~B.}\ \bibnamefont
  {Laughlin}},\ }\bibfield  {title} {\bibinfo {title} {Quantized hall
  conductivity in two dimensions},\ }\href
  {https://doi.org/10.1103/PhysRevB.23.5632} {\bibfield  {journal} {\bibinfo
  {journal} {Phys. Rev. B}\ }\textbf {\bibinfo {volume} {23}},\ \bibinfo
  {pages} {5632} (\bibinfo {year} {1981})}\BibitemShut {NoStop}%
\bibitem [{\citenamefont {Hofstadter}(1976)}]{Hofstadter-PRB}%
  \BibitemOpen
  \bibfield  {author} {\bibinfo {author} {\bibfnamefont {D.~R.}\ \bibnamefont
  {Hofstadter}},\ }\bibfield  {title} {\bibinfo {title} {Energy levels and wave
  functions of bloch electrons in rational and irrational magnetic fields},\
  }\href {https://doi.org/10.1103/PhysRevB.14.2239} {\bibfield  {journal}
  {\bibinfo  {journal} {Phys. Rev. B}\ }\textbf {\bibinfo {volume} {14}},\
  \bibinfo {pages} {2239} (\bibinfo {year} {1976})}\BibitemShut {NoStop}%
\bibitem [{\citenamefont {Harper}(1955)}]{Harper}%
  \BibitemOpen
  \bibfield  {author} {\bibinfo {author} {\bibfnamefont {P.~G.}\ \bibnamefont
  {Harper}},\ }\bibfield  {title} {\bibinfo {title} {Single band motion of
  conduction electrons in a uniform magnetic field},\ }\href@noop {} {\bibfield
   {journal} {\bibinfo  {journal} {Proc. Phys. Soc.}\ }\textbf {\bibinfo
  {volume} {68}},\ \bibinfo {pages} {874} (\bibinfo {year} {1955})}\BibitemShut
  {NoStop}%
\bibitem [{\citenamefont {Goldman}\ \emph {et~al.}(2016)\citenamefont
  {Goldman}, \citenamefont {Budich},\ and\ \citenamefont
  {Zoller}}]{Goldman-Zoller-NaturePhysics}%
  \BibitemOpen
  \bibfield  {author} {\bibinfo {author} {\bibfnamefont {N.}~\bibnamefont
  {Goldman}}, \bibinfo {author} {\bibfnamefont {J.~C.}\ \bibnamefont
  {Budich}},\ and\ \bibinfo {author} {\bibfnamefont {P.}~\bibnamefont
  {Zoller}},\ }\bibfield  {title} {\bibinfo {title} {Topological quantum matter
  with ultracold gases in optical lattices},\ }\href
  {https://doi.org/10.1038/NPHYS3803} {\bibfield  {journal} {\bibinfo
  {journal} {Nature Physics}\ }\textbf {\bibinfo {volume} {12}},\ \bibinfo
  {pages} {639} (\bibinfo {year} {2016})}\BibitemShut {NoStop}%
\bibitem [{\citenamefont {Goldman}\ \emph {et~al.}(2014)\citenamefont
  {Goldman}, \citenamefont {Juzeli\={u}nas}, \citenamefont {Öhberg},\ and\
  \citenamefont {Spielman}}]{Review-Goldman}%
  \BibitemOpen
  \bibfield  {author} {\bibinfo {author} {\bibfnamefont {N.}~\bibnamefont
  {Goldman}}, \bibinfo {author} {\bibfnamefont {G.}~\bibnamefont
  {Juzeli\={u}nas}}, \bibinfo {author} {\bibfnamefont {P.}~\bibnamefont
  {Öhberg}},\ and\ \bibinfo {author} {\bibfnamefont {I.~B.}\ \bibnamefont
  {Spielman}},\ }\bibfield  {title} {\bibinfo {title} {Light-induced gauge
  fields for ultracold atoms},\ }\href@noop {} {\bibfield  {journal} {\bibinfo
  {journal} {Reports on Progress in Physics}\ }\textbf {\bibinfo {volume}
  {77}},\ \bibinfo {pages} {126401} (\bibinfo {year} {2014})}\BibitemShut
  {NoStop}%
\bibitem [{\citenamefont {Jaksch}\ and\ \citenamefont
  {Zoller}(2003)}]{Jaksch-Zoller}%
  \BibitemOpen
  \bibfield  {author} {\bibinfo {author} {\bibfnamefont {D.}~\bibnamefont
  {Jaksch}}\ and\ \bibinfo {author} {\bibfnamefont {P.}~\bibnamefont
  {Zoller}},\ }\bibfield  {title} {\bibinfo {title} {Creation of effective
  magnetic fields in optical lattices: the hofstadter butterfly for cold
  neutral atoms},\ }\href@noop {} {\bibfield  {journal} {\bibinfo  {journal}
  {New Journal of Physics}\ }\textbf {\bibinfo {volume} {5}},\ \bibinfo {pages}
  {56} (\bibinfo {year} {2003})}\BibitemShut {NoStop}%
\bibitem [{\citenamefont {Brooks}\ and\ \citenamefont
  {Das}(2024)}]{Das_Brooks_short}%
  \BibitemOpen
  \bibfield  {author} {\bibinfo {author} {\bibfnamefont {C.}~\bibnamefont
  {Brooks}}\ and\ \bibinfo {author} {\bibfnamefont {K.~K.}\ \bibnamefont
  {Das}},\ }\bibfield  {title} {\bibinfo {title} {Invariant points amidst gauge
  sensitivity in cylindrical and toroidal lattices},\ }\href
  {https://doi.org/10.1103/PhysRevB.109.115404} {\bibfield  {journal} {\bibinfo
   {journal} {Phys. Rev. B}\ }\textbf {\bibinfo {volume} {109}},\ \bibinfo
  {pages} {115404} (\bibinfo {year} {2024})}\BibitemShut {NoStop}%
\bibitem [{\citenamefont {Stone}(2001)}]{Stone-QHE-book}%
  \BibitemOpen
  \bibfield  {author} {\bibinfo {author} {\bibfnamefont {M.}~\bibnamefont
  {Stone}},\ }\href@noop {} {\emph {\bibinfo {title} {Quantum Hall Effect}}},\
  \bibinfo {edition} {1st}\ ed.\ (\bibinfo  {publisher} {World Scientific},\
  \bibinfo {year} {2001})\BibitemShut {NoStop}%
\bibitem [{\citenamefont {Thouless}(1998)}]{Thouless-book}%
  \BibitemOpen
  \bibfield  {author} {\bibinfo {author} {\bibfnamefont {D.~J.}\ \bibnamefont
  {Thouless}},\ }\href@noop {} {\emph {\bibinfo {title} {Topological Quantum
  Numbers in Nonrelativistic Physics}}}\ (\bibinfo  {publisher} {World
  Scientific},\ \bibinfo {address} {Singapore},\ \bibinfo {year}
  {1998})\BibitemShut {NoStop}%
\bibitem [{\citenamefont {Jendrzejewski}\ \emph {et~al.}(2014)\citenamefont
  {Jendrzejewski}, \citenamefont {Eckel}, \citenamefont {Murray}, \citenamefont
  {Lanier}, \citenamefont {Edwards}, \citenamefont {Lobb},\ and\ \citenamefont
  {Campbell}}]{Campbell_resistive-flow}%
  \BibitemOpen
  \bibfield  {author} {\bibinfo {author} {\bibfnamefont {F.}~\bibnamefont
  {Jendrzejewski}}, \bibinfo {author} {\bibfnamefont {S.}~\bibnamefont
  {Eckel}}, \bibinfo {author} {\bibfnamefont {N.}~\bibnamefont {Murray}},
  \bibinfo {author} {\bibfnamefont {C.}~\bibnamefont {Lanier}}, \bibinfo
  {author} {\bibfnamefont {M.}~\bibnamefont {Edwards}}, \bibinfo {author}
  {\bibfnamefont {C.~J.}\ \bibnamefont {Lobb}},\ and\ \bibinfo {author}
  {\bibfnamefont {G.~K.}\ \bibnamefont {Campbell}},\ }\bibfield  {title}
  {\bibinfo {title} {Resistive flow in a weakly interacting bose-einstein
  condensate},\ }\href {https://doi.org/10.1103/PhysRevLett.113.045305}
  {\bibfield  {journal} {\bibinfo  {journal} {Phys. Rev. Lett.}\ }\textbf
  {\bibinfo {volume} {113}},\ \bibinfo {pages} {045305} (\bibinfo {year}
  {2014})}\BibitemShut {NoStop}%
\bibitem [{\citenamefont {Wright}\ \emph {et~al.}(2013)\citenamefont {Wright},
  \citenamefont {Blakestad}, \citenamefont {Lobb}, \citenamefont {Phillips},\
  and\ \citenamefont {Campbell}}]{Phillips_Campbell_superfluid_2013}%
  \BibitemOpen
  \bibfield  {author} {\bibinfo {author} {\bibfnamefont {K.~C.}\ \bibnamefont
  {Wright}}, \bibinfo {author} {\bibfnamefont {R.~B.}\ \bibnamefont
  {Blakestad}}, \bibinfo {author} {\bibfnamefont {C.~J.}\ \bibnamefont {Lobb}},
  \bibinfo {author} {\bibfnamefont {W.~D.}\ \bibnamefont {Phillips}},\ and\
  \bibinfo {author} {\bibfnamefont {G.~K.}\ \bibnamefont {Campbell}},\
  }\bibfield  {title} {\bibinfo {title} {Driving phase slips in a superfluid
  atom circuit with a rotating weak link},\ }\href
  {https://doi.org/10.1103/PhysRevLett.110.025302} {\bibfield  {journal}
  {\bibinfo  {journal} {Phys. Rev. Lett.}\ }\textbf {\bibinfo {volume} {110}},\
  \bibinfo {pages} {025302} (\bibinfo {year} {2013})}\BibitemShut {NoStop}%
\bibitem [{\citenamefont {Eckel}\ \emph {et~al.}(2014)\citenamefont {Eckel},
  \citenamefont {Lee}, \citenamefont {Jendrzejewski}, \citenamefont {Murray},
  \citenamefont {Clark}, \citenamefont {Lobb}, \citenamefont {Phillips},\ and\
  \citenamefont {Campbell}}]{Phillips_Campbell_hysteresis}%
  \BibitemOpen
  \bibfield  {author} {\bibinfo {author} {\bibfnamefont {S.}~\bibnamefont
  {Eckel}}, \bibinfo {author} {\bibfnamefont {J.~G.}\ \bibnamefont {Lee}},
  \bibinfo {author} {\bibfnamefont {F.}~\bibnamefont {Jendrzejewski}}, \bibinfo
  {author} {\bibfnamefont {N.}~\bibnamefont {Murray}}, \bibinfo {author}
  {\bibfnamefont {C.~W.}\ \bibnamefont {Clark}}, \bibinfo {author}
  {\bibfnamefont {C.~J.}\ \bibnamefont {Lobb}}, \bibinfo {author}
  {\bibfnamefont {M.}~\bibnamefont {Phillips}, \bibfnamefont
  {W.~D.and~Edwards}},\ and\ \bibinfo {author} {\bibfnamefont {G.~K.}\
  \bibnamefont {Campbell}},\ }\bibfield  {title} {\bibinfo {title} {Hysteresis
  in a quantized superfluid /`atomtronic/' circuit},\ }\href@noop {} {\bibfield
   {journal} {\bibinfo  {journal} {Nature}\ }\textbf {\bibinfo {volume}
  {506}},\ \bibinfo {pages} {200} (\bibinfo {year} {2014})}\BibitemShut
  {NoStop}%
\bibitem [{\citenamefont {Cai}\ \emph {et~al.}(2022)\citenamefont {Cai},
  \citenamefont {Allman}, \citenamefont {Sabharwal},\ and\ \citenamefont
  {Wright}}]{K-Wright_ring_PRL}%
  \BibitemOpen
  \bibfield  {author} {\bibinfo {author} {\bibfnamefont {Y.}~\bibnamefont
  {Cai}}, \bibinfo {author} {\bibfnamefont {D.~G.}\ \bibnamefont {Allman}},
  \bibinfo {author} {\bibfnamefont {P.}~\bibnamefont {Sabharwal}},\ and\
  \bibinfo {author} {\bibfnamefont {K.~C.}\ \bibnamefont {Wright}},\ }\bibfield
   {title} {\bibinfo {title} {Persistent currents in rings of ultracold
  fermionic atoms},\ }\href {https://doi.org/10.1103/PhysRevLett.128.150401}
  {\bibfield  {journal} {\bibinfo  {journal} {Phys. Rev. Lett.}\ }\textbf
  {\bibinfo {volume} {128}},\ \bibinfo {pages} {150401} (\bibinfo {year}
  {2022})}\BibitemShut {NoStop}%
\bibitem [{\citenamefont {Franke-Arnold}\ \emph {et~al.}(2007)\citenamefont
  {Franke-Arnold}, \citenamefont {Leach}, \citenamefont {Padgett},
  \citenamefont {Lembessis}, \citenamefont {Ellinas}, \citenamefont {Wright},
  \citenamefont {Girkin}, \citenamefont {\"{O}hberg},\ and\ \citenamefont
  {Arnold}}]{Padgett}%
  \BibitemOpen
  \bibfield  {author} {\bibinfo {author} {\bibfnamefont {S.}~\bibnamefont
  {Franke-Arnold}}, \bibinfo {author} {\bibfnamefont {J.}~\bibnamefont
  {Leach}}, \bibinfo {author} {\bibfnamefont {M.~J.}\ \bibnamefont {Padgett}},
  \bibinfo {author} {\bibfnamefont {V.~E.}\ \bibnamefont {Lembessis}}, \bibinfo
  {author} {\bibfnamefont {D.}~\bibnamefont {Ellinas}}, \bibinfo {author}
  {\bibfnamefont {A.~J.}\ \bibnamefont {Wright}}, \bibinfo {author}
  {\bibfnamefont {J.~M.}\ \bibnamefont {Girkin}}, \bibinfo {author}
  {\bibfnamefont {P.}~\bibnamefont {\"{O}hberg}},\ and\ \bibinfo {author}
  {\bibfnamefont {A.~S.}\ \bibnamefont {Arnold}},\ }\bibfield  {title}
  {\bibinfo {title} {Optical ferris wheel for ultracold atoms},\ }\href@noop {}
  {\bibfield  {journal} {\bibinfo  {journal} {Opt. Express}\ }\textbf {\bibinfo
  {volume} {15}},\ \bibinfo {pages} {8619} (\bibinfo {year}
  {2007})}\BibitemShut {NoStop}%
\bibitem [{\citenamefont {Zambrini}\ and\ \citenamefont
  {Barnett}(2007)}]{Zambrini:07}%
  \BibitemOpen
  \bibfield  {author} {\bibinfo {author} {\bibfnamefont {R.}~\bibnamefont
  {Zambrini}}\ and\ \bibinfo {author} {\bibfnamefont {S.~M.}\ \bibnamefont
  {Barnett}},\ }\bibfield  {title} {\bibinfo {title} {Angular momentum of
  multimode and polarization patterns},\ }\href@noop {} {\bibfield  {journal}
  {\bibinfo  {journal} {Opt. Express}\ }\textbf {\bibinfo {volume} {15}},\
  \bibinfo {pages} {15214} (\bibinfo {year} {2007})}\BibitemShut {NoStop}%
\bibitem [{\citenamefont {Lacki}\ \emph {et~al.}(2016)\citenamefont {Lacki},
  \citenamefont {Pichler}, \citenamefont {Sterdyniak}, \citenamefont {Lyras},
  \citenamefont {Lembessis}, \citenamefont {Al-Dossary}, \citenamefont
  {Budich},\ and\ \citenamefont {Zoller}}]{Lacki-cylinder}%
  \BibitemOpen
  \bibfield  {author} {\bibinfo {author} {\bibfnamefont {M.}~\bibnamefont
  {Lacki}}, \bibinfo {author} {\bibfnamefont {H.}~\bibnamefont {Pichler}},
  \bibinfo {author} {\bibfnamefont {A.}~\bibnamefont {Sterdyniak}}, \bibinfo
  {author} {\bibfnamefont {A.}~\bibnamefont {Lyras}}, \bibinfo {author}
  {\bibfnamefont {V.~E.}\ \bibnamefont {Lembessis}}, \bibinfo {author}
  {\bibfnamefont {O.}~\bibnamefont {Al-Dossary}}, \bibinfo {author}
  {\bibfnamefont {J.~C.}\ \bibnamefont {Budich}},\ and\ \bibinfo {author}
  {\bibfnamefont {P.}~\bibnamefont {Zoller}},\ }\bibfield  {title} {\bibinfo
  {title} {Quantum hall physics with cold atoms in cylindrical optical
  lattices},\ }\href {https://doi.org/10.1103/PhysRevA.93.013604} {\bibfield
  {journal} {\bibinfo  {journal} {Phys. Rev. A}\ }\textbf {\bibinfo {volume}
  {93}},\ \bibinfo {pages} {013604} (\bibinfo {year} {2016})}\BibitemShut
  {NoStop}%
\bibitem [{\citenamefont {Budich}\ \emph {et~al.}(2017)\citenamefont {Budich},
  \citenamefont {Elben}, \citenamefont {Lacki}, \citenamefont {Sterdyniak},
  \citenamefont {Baranov},\ and\ \citenamefont {Zoller}}]{Lacki-torus}%
  \BibitemOpen
  \bibfield  {author} {\bibinfo {author} {\bibfnamefont {J.~C.}\ \bibnamefont
  {Budich}}, \bibinfo {author} {\bibfnamefont {A.}~\bibnamefont {Elben}},
  \bibinfo {author} {\bibfnamefont {M.}~\bibnamefont {Lacki}}, \bibinfo
  {author} {\bibfnamefont {A.}~\bibnamefont {Sterdyniak}}, \bibinfo {author}
  {\bibfnamefont {M.~A.}\ \bibnamefont {Baranov}},\ and\ \bibinfo {author}
  {\bibfnamefont {P.}~\bibnamefont {Zoller}},\ }\bibfield  {title} {\bibinfo
  {title} {Coupled atomic wires in a synthetic magnetic field},\ }\href
  {https://doi.org/10.1103/PhysRevA.95.043632} {\bibfield  {journal} {\bibinfo
  {journal} {Phys. Rev. A}\ }\textbf {\bibinfo {volume} {95}},\ \bibinfo
  {pages} {043632} (\bibinfo {year} {2017})}\BibitemShut {NoStop}%
\bibitem [{\citenamefont {Kim}\ \emph {et~al.}(2018)\citenamefont {Kim},
  \citenamefont {Zhu}, \citenamefont {Porto},\ and\ \citenamefont
  {Hafezi}}]{Porto-torus}%
  \BibitemOpen
  \bibfield  {author} {\bibinfo {author} {\bibfnamefont {H.}~\bibnamefont
  {Kim}}, \bibinfo {author} {\bibfnamefont {G.}~\bibnamefont {Zhu}}, \bibinfo
  {author} {\bibfnamefont {J.~V.}\ \bibnamefont {Porto}},\ and\ \bibinfo
  {author} {\bibfnamefont {M.}~\bibnamefont {Hafezi}},\ }\bibfield  {title}
  {\bibinfo {title} {Optical lattice with torus topology},\ }\href
  {https://doi.org/10.1103/PhysRevLett.121.133002} {\bibfield  {journal}
  {\bibinfo  {journal} {Phys. Rev. Lett.}\ }\textbf {\bibinfo {volume} {121}},\
  \bibinfo {pages} {133002} (\bibinfo {year} {2018})}\BibitemShut {NoStop}%
\bibitem [{\citenamefont {Boada}\ \emph {et~al.}(2012)\citenamefont {Boada},
  \citenamefont {Celi}, \citenamefont {Latorre},\ and\ \citenamefont
  {Lewenstein}}]{Boada-PRL}%
  \BibitemOpen
  \bibfield  {author} {\bibinfo {author} {\bibfnamefont {O.}~\bibnamefont
  {Boada}}, \bibinfo {author} {\bibfnamefont {A.}~\bibnamefont {Celi}},
  \bibinfo {author} {\bibfnamefont {J.~I.}\ \bibnamefont {Latorre}},\ and\
  \bibinfo {author} {\bibfnamefont {M.}~\bibnamefont {Lewenstein}},\ }\bibfield
   {title} {\bibinfo {title} {Quantum simulation of an extra dimension},\
  }\href {https://doi.org/10.1103/PhysRevLett.108.133001} {\bibfield  {journal}
  {\bibinfo  {journal} {Phys. Rev. Lett.}\ }\textbf {\bibinfo {volume} {108}},\
  \bibinfo {pages} {133001} (\bibinfo {year} {2012})}\BibitemShut {NoStop}%
\bibitem [{\citenamefont {Celi}\ \emph {et~al.}(2014)\citenamefont {Celi},
  \citenamefont {Massignan}, \citenamefont {Ruseckas}, \citenamefont {Goldman},
  \citenamefont {Spielman}, \citenamefont {Juzeli\ifmmode~\bar{u}\else
  \={u}\fi{}nas},\ and\ \citenamefont {Lewenstein}}]{Spielman-synthetic-dim}%
  \BibitemOpen
  \bibfield  {author} {\bibinfo {author} {\bibfnamefont {A.}~\bibnamefont
  {Celi}}, \bibinfo {author} {\bibfnamefont {P.}~\bibnamefont {Massignan}},
  \bibinfo {author} {\bibfnamefont {J.}~\bibnamefont {Ruseckas}}, \bibinfo
  {author} {\bibfnamefont {N.}~\bibnamefont {Goldman}}, \bibinfo {author}
  {\bibfnamefont {I.~B.}\ \bibnamefont {Spielman}}, \bibinfo {author}
  {\bibfnamefont {G.}~\bibnamefont {Juzeli\ifmmode~\bar{u}\else
  \={u}\fi{}nas}},\ and\ \bibinfo {author} {\bibfnamefont {M.}~\bibnamefont
  {Lewenstein}},\ }\bibfield  {title} {\bibinfo {title} {Synthetic gauge fields
  in synthetic dimensions},\ }\href
  {https://doi.org/10.1103/PhysRevLett.112.043001} {\bibfield  {journal}
  {\bibinfo  {journal} {Phys. Rev. Lett.}\ }\textbf {\bibinfo {volume} {112}},\
  \bibinfo {pages} {043001} (\bibinfo {year} {2014})}\BibitemShut {NoStop}%
\bibitem [{\citenamefont {Zhang}\ \emph {et~al.}(2021)\citenamefont {Zhang},
  \citenamefont {Yan},\ and\ \citenamefont {Zhou}}]{Zhou_Hall_loc_PRL}%
  \BibitemOpen
  \bibfield  {author} {\bibinfo {author} {\bibfnamefont {R.}~\bibnamefont
  {Zhang}}, \bibinfo {author} {\bibfnamefont {Y.}~\bibnamefont {Yan}},\ and\
  \bibinfo {author} {\bibfnamefont {Q.}~\bibnamefont {Zhou}},\ }\bibfield
  {title} {\bibinfo {title} {Localization on a synthetic hall cylinder},\
  }\href {https://doi.org/10.1103/PhysRevLett.126.193001} {\bibfield  {journal}
  {\bibinfo  {journal} {Phys. Rev. Lett.}\ }\textbf {\bibinfo {volume} {126}},\
  \bibinfo {pages} {193001} (\bibinfo {year} {2021})}\BibitemShut {NoStop}%
\bibitem [{\citenamefont {Yan}\ \emph {et~al.}(2019)\citenamefont {Yan},
  \citenamefont {Zhang}, \citenamefont {Choudhury},\ and\ \citenamefont
  {Zhou}}]{Zhou_PRL_cylinder_tori}%
  \BibitemOpen
  \bibfield  {author} {\bibinfo {author} {\bibfnamefont {Y.}~\bibnamefont
  {Yan}}, \bibinfo {author} {\bibfnamefont {S.-L.}\ \bibnamefont {Zhang}},
  \bibinfo {author} {\bibfnamefont {S.}~\bibnamefont {Choudhury}},\ and\
  \bibinfo {author} {\bibfnamefont {Q.}~\bibnamefont {Zhou}},\ }\bibfield
  {title} {\bibinfo {title} {Emergent periodic and quasiperiodic lattices on
  surfaces of synthetic hall tori and synthetic hall cylinders},\ }\href
  {https://doi.org/10.1103/PhysRevLett.123.260405} {\bibfield  {journal}
  {\bibinfo  {journal} {Phys. Rev. Lett.}\ }\textbf {\bibinfo {volume} {123}},\
  \bibinfo {pages} {260405} (\bibinfo {year} {2019})}\BibitemShut {NoStop}%
\bibitem [{\citenamefont {Liang}\ \emph {et~al.}(2021)\citenamefont {Liang},
  \citenamefont {Trypogeorgos}, \citenamefont {Vald\'es-Curiel}, \citenamefont
  {Tao}, \citenamefont {Zhao},\ and\ \citenamefont
  {Spielman}}]{Spielman-synthetic-cyl}%
  \BibitemOpen
  \bibfield  {author} {\bibinfo {author} {\bibfnamefont {Q.-Y.}\ \bibnamefont
  {Liang}}, \bibinfo {author} {\bibfnamefont {D.}~\bibnamefont {Trypogeorgos}},
  \bibinfo {author} {\bibfnamefont {A.}~\bibnamefont {Vald\'es-Curiel}},
  \bibinfo {author} {\bibfnamefont {J.}~\bibnamefont {Tao}}, \bibinfo {author}
  {\bibfnamefont {M.}~\bibnamefont {Zhao}},\ and\ \bibinfo {author}
  {\bibfnamefont {I.~B.}\ \bibnamefont {Spielman}},\ }\bibfield  {title}
  {\bibinfo {title} {Coherence and decoherence in the harper-hofstadter
  model},\ }\href {https://doi.org/10.1103/PhysRevResearch.3.023058} {\bibfield
   {journal} {\bibinfo  {journal} {Phys. Rev. Res.}\ }\textbf {\bibinfo
  {volume} {3}},\ \bibinfo {pages} {023058} (\bibinfo {year}
  {2021})}\BibitemShut {NoStop}%
\bibitem [{\citenamefont {Li}\ \emph {et~al.}(2022)\citenamefont {Li},
  \citenamefont {Yan}, \citenamefont {Feng}, \citenamefont {Choudhury},
  \citenamefont {Blasing}, \citenamefont {Zhou},\ and\ \citenamefont
  {Chen}}]{Zhou_PRX_Hall_cylinder}%
  \BibitemOpen
  \bibfield  {author} {\bibinfo {author} {\bibfnamefont {C.-H.}\ \bibnamefont
  {Li}}, \bibinfo {author} {\bibfnamefont {Y.}~\bibnamefont {Yan}}, \bibinfo
  {author} {\bibfnamefont {S.-W.}\ \bibnamefont {Feng}}, \bibinfo {author}
  {\bibfnamefont {S.}~\bibnamefont {Choudhury}}, \bibinfo {author}
  {\bibfnamefont {D.~B.}\ \bibnamefont {Blasing}}, \bibinfo {author}
  {\bibfnamefont {Q.}~\bibnamefont {Zhou}},\ and\ \bibinfo {author}
  {\bibfnamefont {Y.~P.}\ \bibnamefont {Chen}},\ }\bibfield  {title} {\bibinfo
  {title} {Bose-einstein condensate on a synthetic topological hall cylinder},\
  }\href {https://doi.org/10.1103/PRXQuantum.3.010316} {\bibfield  {journal}
  {\bibinfo  {journal} {PRX Quantum}\ }\textbf {\bibinfo {volume} {3}},\
  \bibinfo {pages} {010316} (\bibinfo {year} {2022})}\BibitemShut {NoStop}%
\bibitem [{\citenamefont {Fabre}\ \emph {et~al.}(2022)\citenamefont {Fabre},
  \citenamefont {Bouhiron}, \citenamefont {Satoor}, \citenamefont {Lopes},\
  and\ \citenamefont {Nascimbene}}]{Fabre_Laughlin_expt}%
  \BibitemOpen
  \bibfield  {author} {\bibinfo {author} {\bibfnamefont {A.}~\bibnamefont
  {Fabre}}, \bibinfo {author} {\bibfnamefont {J.-B.}\ \bibnamefont {Bouhiron}},
  \bibinfo {author} {\bibfnamefont {T.}~\bibnamefont {Satoor}}, \bibinfo
  {author} {\bibfnamefont {R.}~\bibnamefont {Lopes}},\ and\ \bibinfo {author}
  {\bibfnamefont {S.}~\bibnamefont {Nascimbene}},\ }\bibfield  {title}
  {\bibinfo {title} {Laughlin's topological charge pump in an atomic hall
  cylinder},\ }\href {https://doi.org/10.1103/PhysRevLett.128.173202}
  {\bibfield  {journal} {\bibinfo  {journal} {Phys. Rev. Lett.}\ }\textbf
  {\bibinfo {volume} {128}},\ \bibinfo {pages} {173202} (\bibinfo {year}
  {2022})}\BibitemShut {NoStop}%
\bibitem [{\citenamefont {Barbarino}\ \emph {et~al.}(2018)\citenamefont
  {Barbarino}, \citenamefont {Dalmonte}, \citenamefont {Fazio},\ and\
  \citenamefont {Santoro}}]{Santoro-synthetic-Hall}%
  \BibitemOpen
  \bibfield  {author} {\bibinfo {author} {\bibfnamefont {S.}~\bibnamefont
  {Barbarino}}, \bibinfo {author} {\bibfnamefont {M.}~\bibnamefont {Dalmonte}},
  \bibinfo {author} {\bibfnamefont {R.}~\bibnamefont {Fazio}},\ and\ \bibinfo
  {author} {\bibfnamefont {G.~E.}\ \bibnamefont {Santoro}},\ }\bibfield
  {title} {\bibinfo {title} {Topological phases in frustrated synthetic ladders
  with an odd number of legs},\ }\href
  {https://doi.org/10.1103/PhysRevA.97.013634} {\bibfield  {journal} {\bibinfo
  {journal} {Phys. Rev. A}\ }\textbf {\bibinfo {volume} {97}},\ \bibinfo
  {pages} {013634} (\bibinfo {year} {2018})}\BibitemShut {NoStop}%
\bibitem [{\citenamefont {Han}\ \emph {et~al.}(2019)\citenamefont {Han},
  \citenamefont {Kang},\ and\ \citenamefont {Shin}}]{Shin_bandgap-closing-PRL}%
  \BibitemOpen
  \bibfield  {author} {\bibinfo {author} {\bibfnamefont {J.~H.}\ \bibnamefont
  {Han}}, \bibinfo {author} {\bibfnamefont {J.~H.}\ \bibnamefont {Kang}},\ and\
  \bibinfo {author} {\bibfnamefont {Y.}~\bibnamefont {Shin}},\ }\bibfield
  {title} {\bibinfo {title} {Band gap closing in a synthetic hall tube of
  neutral fermions},\ }\href {https://doi.org/10.1103/PhysRevLett.122.065303}
  {\bibfield  {journal} {\bibinfo  {journal} {Phys. Rev. Lett.}\ }\textbf
  {\bibinfo {volume} {122}},\ \bibinfo {pages} {065303} (\bibinfo {year}
  {2019})}\BibitemShut {NoStop}%
\bibitem [{\citenamefont {Han}\ \emph {et~al.}(2022)\citenamefont {Han},
  \citenamefont {Bae},\ and\ \citenamefont {Shin}}]{Shin-synthetic-Hall}%
  \BibitemOpen
  \bibfield  {author} {\bibinfo {author} {\bibfnamefont {J.~H.}\ \bibnamefont
  {Han}}, \bibinfo {author} {\bibfnamefont {D.}~\bibnamefont {Bae}},\ and\
  \bibinfo {author} {\bibfnamefont {Y.}~\bibnamefont {Shin}},\ }\bibfield
  {title} {\bibinfo {title} {Synthetic hall ladder with tunable magnetic
  flux},\ }\href {https://doi.org/10.1103/PhysRevA.105.043306} {\bibfield
  {journal} {\bibinfo  {journal} {Phys. Rev. A}\ }\textbf {\bibinfo {volume}
  {105}},\ \bibinfo {pages} {043306} (\bibinfo {year} {2022})}\BibitemShut
  {NoStop}%
\bibitem [{\citenamefont {Aidelsburger}\ \emph {et~al.}(2014)\citenamefont
  {Aidelsburger}, \citenamefont {Lohse}, \citenamefont {Schweizer},
  \citenamefont {Atala}, \citenamefont {Barreiro}, \citenamefont {Nascimbene},
  \citenamefont {Cooper}, \citenamefont {Bloch},\ and\ \citenamefont
  {Goldman}}]{A5}%
  \BibitemOpen
  \bibfield  {author} {\bibinfo {author} {\bibfnamefont {M.}~\bibnamefont
  {Aidelsburger}}, \bibinfo {author} {\bibnamefont {Lohse}}, \bibinfo {author}
  {\bibfnamefont {C.}~\bibnamefont {Schweizer}}, \bibinfo {author}
  {\bibfnamefont {M.}~\bibnamefont {Atala}, \bibfnamefont {M.}}, \bibinfo
  {author} {\bibfnamefont {J.~T.}\ \bibnamefont {Barreiro}}, \bibinfo {author}
  {\bibfnamefont {S.}~\bibnamefont {Nascimbene}}, \bibinfo {author}
  {\bibfnamefont {N.~R.}\ \bibnamefont {Cooper}}, \bibinfo {author}
  {\bibfnamefont {I.}~\bibnamefont {Bloch}},\ and\ \bibinfo {author}
  {\bibfnamefont {N.}~\bibnamefont {Goldman}},\ }\bibfield  {title} {\bibinfo
  {title} {Measuring the chern number of hofstadter bands with ultracold
  bosonic atoms},\ }\href {https://doi.org/10.1038/nphys3171} {\bibfield
  {journal} {\bibinfo  {journal} {Nature Physics}\ }\textbf {\bibinfo {volume}
  {162}},\ \bibinfo {pages} {11} (\bibinfo {year} {2014})}\BibitemShut
  {NoStop}%
\bibitem [{\citenamefont {Stuhl}\ \emph {et~al.}(2015)\citenamefont {Stuhl},
  \citenamefont {Lu}, \citenamefont {Aycock}, \citenamefont {Genkina},\ and\
  \citenamefont {Spielman}}]{A2}%
  \BibitemOpen
  \bibfield  {author} {\bibinfo {author} {\bibfnamefont {B.~K.}\ \bibnamefont
  {Stuhl}}, \bibinfo {author} {\bibfnamefont {H.-I.}\ \bibnamefont {Lu}},
  \bibinfo {author} {\bibfnamefont {L.~M.}\ \bibnamefont {Aycock}}, \bibinfo
  {author} {\bibfnamefont {D.}~\bibnamefont {Genkina}},\ and\ \bibinfo {author}
  {\bibfnamefont {I.~B.}\ \bibnamefont {Spielman}},\ }\bibfield  {title}
  {\bibinfo {title} {Visualizing edge states with an atomic bose gas in the
  quantum hall regime},\ }\href {https://doi.org/10.1126/science.aaa8515}
  {\bibfield  {journal} {\bibinfo  {journal} {Science}\ }\textbf {\bibinfo
  {volume} {349}},\ \bibinfo {pages} {1514} (\bibinfo {year}
  {2015})}\BibitemShut {NoStop}%
\bibitem [{\citenamefont {Lelas}\ \emph {et~al.}(2016)\citenamefont {Lelas},
  \citenamefont {Drpi\'{c}}, \citenamefont {Dub\v{c}ek}, \citenamefont
  {Juki\'{c}}, \citenamefont {Pezer},\ and\ \citenamefont
  {Buljan}}]{Lelas_2016}%
  \BibitemOpen
  \bibfield  {author} {\bibinfo {author} {\bibfnamefont {K.}~\bibnamefont
  {Lelas}}, \bibinfo {author} {\bibfnamefont {N.}~\bibnamefont {Drpi\'{c}}},
  \bibinfo {author} {\bibfnamefont {T.}~\bibnamefont {Dub\v{c}ek}}, \bibinfo
  {author} {\bibfnamefont {D.}~\bibnamefont {Juki\'{c}}}, \bibinfo {author}
  {\bibfnamefont {R.}~\bibnamefont {Pezer}},\ and\ \bibinfo {author}
  {\bibfnamefont {H.}~\bibnamefont {Buljan}},\ }\bibfield  {title} {\bibinfo
  {title} {Laser assisted tunneling in a tonks-girardeau gas},\ }\href@noop {}
  {\bibfield  {journal} {\bibinfo  {journal} {New Journal of Physics}\ }\textbf
  {\bibinfo {volume} {18}},\ \bibinfo {pages} {095002} (\bibinfo {year}
  {2016})}\BibitemShut {NoStop}%
\bibitem [{\citenamefont {Rosson}\ \emph {et~al.}(2019)\citenamefont {Rosson},
  \citenamefont {Lubasch}, \citenamefont {Kiffner},\ and\ \citenamefont
  {Jaksch}}]{A10}%
  \BibitemOpen
  \bibfield  {author} {\bibinfo {author} {\bibfnamefont {P.}~\bibnamefont
  {Rosson}}, \bibinfo {author} {\bibfnamefont {M.}~\bibnamefont {Lubasch}},
  \bibinfo {author} {\bibfnamefont {M.}~\bibnamefont {Kiffner}},\ and\ \bibinfo
  {author} {\bibfnamefont {D.}~\bibnamefont {Jaksch}},\ }\bibfield  {title}
  {\bibinfo {title} {Bosonic fractional quantum hall states on a finite
  cylinder},\ }\href {https://doi.org/10.1103/PhysRevA.99.033603} {\bibfield
  {journal} {\bibinfo  {journal} {Phys. Rev. A}\ }\textbf {\bibinfo {volume}
  {99}},\ \bibinfo {pages} {033603} (\bibinfo {year} {2019})}\BibitemShut
  {NoStop}%
\bibitem [{\citenamefont {Zak}(1964)}]{Zak-1}%
  \BibitemOpen
  \bibfield  {author} {\bibinfo {author} {\bibfnamefont {J.}~\bibnamefont
  {Zak}},\ }\bibfield  {title} {\bibinfo {title} {Magnetic translation group},\
  }\href {https://doi.org/10.1103/PhysRev.134.A1602} {\bibfield  {journal}
  {\bibinfo  {journal} {Phys. Rev.}\ }\textbf {\bibinfo {volume} {134}},\
  \bibinfo {pages} {A1602} (\bibinfo {year} {1964})}\BibitemShut {NoStop}%
\bibitem [{\citenamefont {Halperin}(1982)}]{Halperin-QHE}%
  \BibitemOpen
  \bibfield  {author} {\bibinfo {author} {\bibfnamefont {B.~I.}\ \bibnamefont
  {Halperin}},\ }\bibfield  {title} {\bibinfo {title} {Quantized hall
  conductance, current-carrying edge states, and the existence of extended
  states in a two-dimensional disordered potential},\ }\href
  {https://doi.org/10.1103/PhysRevB.25.2185} {\bibfield  {journal} {\bibinfo
  {journal} {Phys. Rev. B}\ }\textbf {\bibinfo {volume} {25}},\ \bibinfo
  {pages} {2185} (\bibinfo {year} {1982})}\BibitemShut {NoStop}%
\bibitem [{\citenamefont {Fukui}\ \emph {et~al.}(2005)\citenamefont {Fukui},
  \citenamefont {Hatsugai},\ and\ \citenamefont {Suzuki}}]{Suzuki-Chern}%
  \BibitemOpen
  \bibfield  {author} {\bibinfo {author} {\bibfnamefont {T.}~\bibnamefont
  {Fukui}}, \bibinfo {author} {\bibfnamefont {Y.}~\bibnamefont {Hatsugai}},\
  and\ \bibinfo {author} {\bibfnamefont {H.}~\bibnamefont {Suzuki}},\
  }\bibfield  {title} {\bibinfo {title} {Chern numbers in discretized brillouin
  zone: efficient method of computing (spin) hall conductances},\ }\href@noop
  {} {\bibfield  {journal} {\bibinfo  {journal} {J. Phys. Soc. Jpn.}\ }\textbf
  {\bibinfo {volume} {74}},\ \bibinfo {pages} {1674} (\bibinfo {year}
  {2005})}\BibitemShut {NoStop}%
\bibitem [{\citenamefont {Lin}\ \emph {et~al.}(2023)\citenamefont {Lin},
  \citenamefont {Ke}, \citenamefont {Zhang},\ and\ \citenamefont
  {Lee}}]{Chern-1}%
  \BibitemOpen
  \bibfield  {author} {\bibinfo {author} {\bibfnamefont {L.}~\bibnamefont
  {Lin}}, \bibinfo {author} {\bibfnamefont {Y.}~\bibnamefont {Ke}}, \bibinfo
  {author} {\bibfnamefont {L.}~\bibnamefont {Zhang}},\ and\ \bibinfo {author}
  {\bibfnamefont {C.}~\bibnamefont {Lee}},\ }\bibfield  {title} {\bibinfo
  {title} {Calculations of chern number: equivalence of real-space and
  twisted-boundary-condition formulae},\ }\href@noop {} {\bibfield  {journal}
  {\bibinfo  {journal} {arXiv:2308.04164 [quant-ph]}\ } (\bibinfo {year}
  {2023})}\BibitemShut {NoStop}%
\bibitem [{\citenamefont {Das}(2020)}]{Das-PRL-localization}%
  \BibitemOpen
  \bibfield  {author} {\bibinfo {author} {\bibfnamefont {K.~K.}\ \bibnamefont
  {Das}},\ }\bibfield  {title} {\bibinfo {title} {Significance and sensor
  utility of phase in quantum localization transition},\ }\href
  {https://doi.org/10.1103/PhysRevLett.125.070401} {\bibfield  {journal}
  {\bibinfo  {journal} {Phys. Rev. Lett.}\ }\textbf {\bibinfo {volume} {125}},\
  \bibinfo {pages} {070401} (\bibinfo {year} {2020})}\BibitemShut {NoStop}%
\end{thebibliography}

%

\end{document}